\documentclass[prb,twocolumn,amsmath,amssymb]{revtex4-1}

\usepackage{amsthm}
\usepackage{color}
\usepackage{graphicx}
\usepackage{dcolumn}
\usepackage{bm}
\usepackage{comment}
\usepackage{xr}
\newcommand{\vk}{{\mbox{\boldmath$k$}}}

\renewcommand{\paragraph}[1]{{\par\it #1.---}\ignorespaces}

\newtheorem*{theorem*}{Theorem}
\newtheorem*{prop*}{Proposition}

\begin{document}


\title{Global band topology of simple and double Dirac-point (semi-)metals}

\author{Adrien Bouhon}%
\email{adrien.bouhon@physics.uu.se}
\author{Annica M.~Black-Schaffer}
\affiliation{%
Department of Physics and Astronomy, Uppsala University, Box 516, SE-751 21 Uppsala, Sweden 
}%

\date{\today}

\begin{abstract}
We combine space group representation theory together with scanning of closed subdomains of the Brillouin zone with Wilson loops to algebraically determine global band structure topology. Considering space group $\#$19 as a case study, we show that the energy ordering of the irreducible representations at the high-symmetry points $\{\Gamma,S,T,U\}$ fully determines the global band topology, with all topological classes characterized through their simple and double Dirac-points. 
\end{abstract}

\maketitle

Topological (semi-)metals with their protected band-crossing Dirac (or Weyl) points have recently attracted a lot of interest. Like many other topological phases of matter, symmetries play an important role in the understanding, classification, and prediction of topological semimetals \cite{Class_sym_review}. While much work have focused on local characterization of symmetry protected Dirac points \cite{YoungKane_simple, YangNagaosa, ChiuSchnyder_reflection, YMFurusaki, WiederKane_double,Sato_1} and also on specific global features \cite{Kane_nonsym,Kane_line,Sato_2,Sato_3,Furusaki_line,Kane_2D_SOC,Thomas_line,ZhaoSchnyder_1D,1611.07925,GB_line}, a systematic treatment of the global band topology enforced by the crystal space group is still missing.

Space group representation theory fully determines the symmetry-protected band-crossings occurring at high-symmetry points or lines of the Brillouin zone (BZ), each being treated separately \cite{BradCrack}.
It was also early realized that nonsymmorphic space groups can realize \textit{connected elementary energy bands} \cite{Zak1,Zak2,Zak3,Zak4}, i.e.~a minimum number of bands that are connected through enough contact points such that one can travel continuously through these bands over the BZ. 
This leads to an additional global type of symmetry protected Dirac points that can be moved in some determined regions of the BZ but are globally unavoidable. 
A consequence for such space groups is the tightening of the necessary electron filling condition for realizing a band insulator. While the filling number must usually only be even for a insulating state, it typically needs to be within a proper subset of this for a nonsymmorphic space group \cite{Vishwanath1}. Conversely, it is sufficient (but not necessary) to violate this tighter filling condition to achieve a (semi-)metallic phase. 

Still, one main question has remained open so far: For a given space group $\mathcal{G}$, what is the global topology of the band structure including all Dirac points, which in turn also provides the filling condition for a topological (semi-)metallic state?
In this work we show that combining space group representation theory together with Wilson loop techniques to calculate the Berry phase leads to a definitive answer to the question, using space group $\#$19 (SG19) as a case study \footnote{In this work we neglect spin-orbit coupling and use a fully spin-polarized description, but otherwise assume time-reversal symmetry. A brief discussion of breaking TRS is included in the Supplementary materials \cite{SM}.}. In fact, we show that the global band topology, including all symmetry protected Dirac points, is fully determined simply by the energy ordering of the irreducible representations (IRREPs) at the high-symmetry points $\{\Gamma,S,T,U\}$.

\begin{figure}[t]
\centering
\begin{tabular}{c} 
	\includegraphics[width=0.95\linewidth]{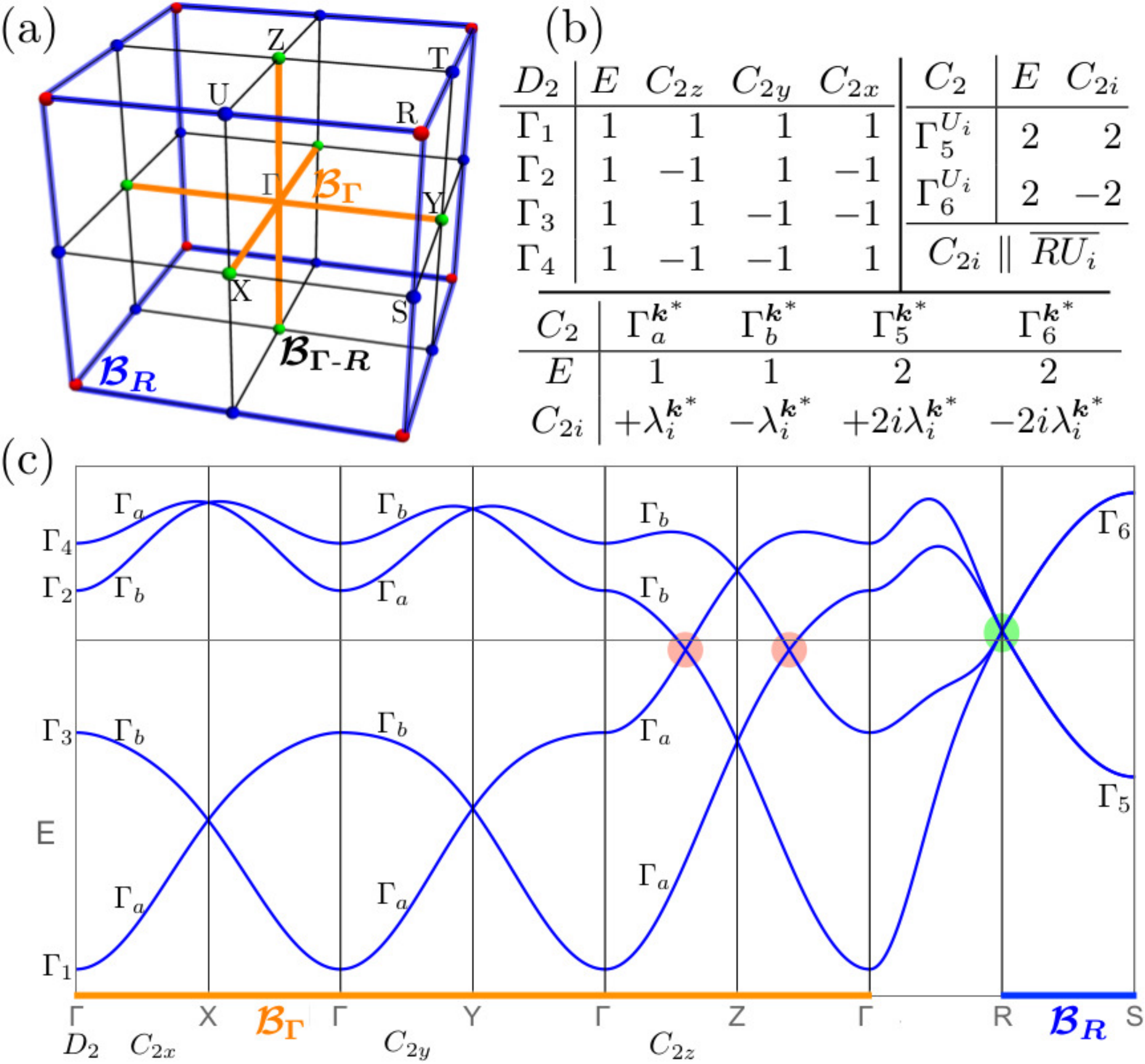} 
\end{tabular}
\caption{\label{fig_SG19} (a) BZ with high-symmetry points and lines of SG19. (b) Character tables of the IRREPS for the point groups $D_2$ and $C_2$ with $\lambda^{\boldsymbol{k}^*}_{i} = \mathrm{e}^{-i \boldsymbol{k}^*\cdot \boldsymbol{\tau}_i}$, where $\boldsymbol{k}^*$ belongs to one line of $\mathcal{B}_{\Gamma}$ for $\{\Gamma_a,\Gamma_b\}$ and $\mathcal{B}_{R}$ for $\{\Gamma_5,\Gamma_6\}$. (c) Electronic band structure of a four-band tight-binding model in SG19. Unavoidable globally protected simple Dirac points ($\vert C_1\vert=1$) in red and double Dirac point ($\vert C_1\vert=2$) in green.
}
\end{figure}

\paragraph{$4N$-band structures in SG19}
The nonsymmorphic SG19 (P$2_12_12_1$) is composed of a primitive orthorhombic Bravais lattice and three screw axes $\{g\vert \boldsymbol{\tau}_g\}$ with the point group elements $g\in D_2 = \{E,C_{2z},C_{2y},C_{2x}\}$ and the fractional translations $ \boldsymbol{\tau}_x = (\boldsymbol{a}_1+\boldsymbol{a}_2)/2$, $ \boldsymbol{\tau}_y = (\boldsymbol{a}_2+\boldsymbol{a}_3)/2$, $ \boldsymbol{\tau}_z = (\boldsymbol{a}_1+\boldsymbol{a}_3)/2$, where $\{\boldsymbol{a}_i\}_{i=1,2,3}$ are the primitive lattice vectors.
Since SG19 has a single Wyckoff position with no symmetry, the set of all (one-dimensional, 1D) IRREPs at $\Gamma$ must split into $N$ copies of the four IRREPs of $D_2$, $\{\Gamma_1^{\boldsymbol{0}}, \Gamma_2^{\boldsymbol{0}},\Gamma_3^{\boldsymbol{0}},\Gamma_4^{\boldsymbol{0}}\}$ defined by the character table in Fig.~\ref{fig_SG19}(b)~\footnote{Each lattice site has three inequivalent partners under screw symmetries with which they form a basis for each of the four IRREPs of $D_2$, see SM \cite{SM}.}. Likewise, the set of all 2D IRREPs at the points $U_i \in \{S,T,U\}$ splits into $N$ copies of the two projective IRREPs $\{\Gamma^{U_i}_5,\Gamma^{U_i}_6\}$, also given in Fig.~\ref{fig_SG19}(b) \cite{BradCrack, Bilbao}. 
It is convenient to split the high-symmetry lines into three distinct BZ subspaces: $\mathcal{B}_{\Gamma}=\bigcup_{X_i=X,Y,Z} \overline{\Gamma X_i}$ (high-symmetry lines crossing $\Gamma$), $\mathcal{B}_{R}=\bigcup_{U_i} \overline{R U_i}$ (high-symmetry lines crossing $R$), and $\mathcal{B}_{\Gamma\text{-}R} = \bigcup_{X_i,U_i}\overline{X_i U_i}$ (high-symmetry lines connecting $\mathcal{B}_{\Gamma}$ and $\mathcal{B}_{R}$), see Fig.~\ref{fig_SG19}(a).
Only a unique 2D IRREP is allowed on $\mathcal{B}_{\Gamma\text{-}R}$ \cite{BradCrack,Bilbao}. This leads to $\mathcal{B}_{\Gamma}$ and $\mathcal{B}_{R}$ being \textit{symmetry independent} since their IRREPs are not constrained by their compatibility relation into $\mathcal{B}_{\Gamma\text{-}R}$.

With this background we directly state our first main result:
{\it Any $4N$-band structure of SG19 can be reconstructed by hand from the list of energy ordered IRREPs at the high-symmetry points $\{\Gamma,S,T,U\}$ by applying Rules 1-4 below.}
To show this we start with the $\mathcal{B}_{\Gamma}$ subspace. Labeling all the energy eigenvalues at $\Gamma$ according to their band index ($n=1,\dots,N$) and IRREP ($j=1,2,3,4$) as $E^{n}_{j}(\boldsymbol{0})$, we can follow smoothly each eigenvalue branch over $\mathcal{B}_{\Gamma}$ (we define this as the \textit{smooth gauge}, see Supplementary material (SM) \cite{SM}). We write $\mathcal{P}_{\overline{\Gamma\boldsymbol{b}_i}} = (j_1 j_2) (j_3 j_4)$ for the two-by-two permutation in energy of the four bands $\{E^{n_1}_{j_1},E^{n_2}_{j_2},E^{n_3}_{j_3},E^{n_4}_{j_4}\}$ under a shift by a primitive reciprocal lattice vector $\boldsymbol{b}_i$ ($\boldsymbol{b}_i \parallel \overline{\Gamma X_i}$). 
We then get (see SM \cite{SM}) {\it Rule 1: All bands are permuted along each line of $\mathcal{B}_{\Gamma}$ according to Table~\ref{permutation_SG19}.} 
\begin{table}[h]
 \begin{tabular}{|c|c|c||c|} 
 \hline 
  $\mathcal{P}_{\overline{\Gamma\boldsymbol{b}_1}}$  & $\mathcal{P}_{\overline{\Gamma\boldsymbol{b}_2}}$ & $\mathcal{P}_{\overline{\Gamma\boldsymbol{b}_3}}$ & $\mathcal{P}_{\overline{S\boldsymbol{b}_3}},\mathcal{P}_{\overline{T\boldsymbol{b}_1}},\mathcal{P}_{\overline{U\boldsymbol{b}_2}}$ \\
 \hline
 \hline
 $\begin{array}{c}   (12)(34) \\  (13)(24) \end{array} $ & 
 		$\begin{array}{c}  (13)(24) \\  (14)(23)  \end{array}$ & 
		$\begin{array}{c}  (12)(34) \\ (14)(23)  \end{array}$  & (56) \\
 \hline
 \end{tabular}
\caption{\label{permutation_SG19} Band permutation rules in $\mathcal{B}_{\Gamma}$ and $\mathcal{B}_R$. Bands are labeled according to their IRREPs at $\Gamma$ for $\mathcal{B}_{\Gamma}$ and $U_i$ for $\mathcal{B}_R$. }
\end{table}
Permutations only happen between bands belonging to different IRREPs along each line $\overline{\Gamma X_i}$, i.e. with different compatibility relations $\Gamma^{\boldsymbol{0}}_{j} \rightarrow \Gamma^{\boldsymbol{k}^*\in\overline{\Gamma X_i}}_{k}$, $k=a,b$ (see Fig.~\ref{fig_SG19}(b)). Hence, any two permuted bands have a symmetry protected crossing. In fact, these crossing points are at the middle points $\{X,Y,Z\}$ because of their $D_2$ symmetry \cite{BradCrack,Bilbao}~\footnote{Note that these crossings are actually part of twofold degenerate lines over the whole $\mathcal{B}_{\Gamma\text{-}R}$ subspace.}. 
In addition, we have {\it Rule 2: For any two bands $E^{n_1}_{j_1}$ and $E^{n_2}_{j_2}$ at $\Gamma$ with the same compatibility relation into $\overline{\Gamma X_i} $, we have} 
\begin{align}
E^{n_1}_{j_1}(\boldsymbol{0}) \gtrless E^{n_2}_{j_2}(\boldsymbol{0}) \Leftrightarrow E^{n_1}_{j_1}(\boldsymbol{k}^*) \gtrless E^{n_2}_{j_2}(\boldsymbol{k}^*), \forall~\boldsymbol{k}^* \in \overline{\Gamma X_i} \nonumber\;.
\end{align}
This rule is a straightforward consequence of (i) smoothness of the eigenvalues as functions of $\vk$ and (ii) two states with the same compatibility relation into a given $\boldsymbol{k}^*$ being able to hybridize, hence forbidding symmetry protected band-crossings.
Applying Rules 1--2 we readily conclude that any isolated four-band subspace (i.e.~separated by an energy gap)  realizes two distinct permutations of Table \ref{permutation_SG19} over $\mathcal{B}_{\Gamma}$, no more no less. This leads to two new Dirac points (apart from the crossings at $X_i$) somewhere along one of the lines $\{\overline{\Gamma X_i}\}$, with $i$ determined only by the order of the IRREPs at $\Gamma$. These two Dirac points are protected by the global band topology. In Fig.~\ref{fig_SG19}(c) we provide a four-band tight-binding example with these two Dirac points (red)  on $\overline{\Gamma Z}$. 

Next we consider the $\mathcal{B}_{R}$ subspace. Labeling the bands at $U_i$ as $E^{n}_{l}(U_i)$, with $l=5,6$, and again assuming the smooth gauge, we write as $\mathcal{P}_{\overline{U_i\boldsymbol{b}_j}} = (l_1 l_2)$ the two-by-two permutation in energy of the two bands $\{E^{n_1}_{l_1},E^{n_2}_{l_2}\}$ under a shift from $U_i$ by $\boldsymbol{b}_{j} \parallel \overline{RU_i}$. We then get
{\it Rule 3: All the bands are permuted along each line of $\mathcal{B}_{R}$ according to $\mathcal{P}_{\overline{S\boldsymbol{b}_3}} = \mathcal{P}_{\overline{T\boldsymbol{b}_1}} = \mathcal{P}_{\overline{U\boldsymbol{b}_2}} = (56).$}
These band permutations enforces one symmetry protected Dirac point along each line $\overline{R U_i}$, since the two bands $\{ 5,6\}$ correspond to  distinct IRREPs on these lines, see Fig.~\ref{fig_SG19}(b). Because of the $D_2$ symmetry of the midpoint $R$, these crossings will always be at $R$, leading to a fourfold degeneracy as indicated in green in Fig.~\ref{fig_SG19}(c). Similarly to Rule 2, we finally have
{\it Rule 4: For any two bands $E^{n_1}_{l_1}$ and $E^{n_2}_{l_2}$ at $U_i$ with the same IRREPs, i.e.~$l_1=l_2$, we have
\begin{align*}
	E^{n_1}_{l_1}(U_i) \gtrless E^{n_2}_{l_1}(U_i) \Leftrightarrow E^{n_1}_{l_1}(\boldsymbol{k}^*) \gtrless E^{n_2}_{l_1}(\boldsymbol{k}^*), \forall~\boldsymbol{k}^* \in \overline{RU_i}.
\end{align*}}Together, Rules 3--4 fully determine global band structure in the $\mathcal{B}_{R}$ subspace.

Left to consider is the $\mathcal{B}_{\Gamma\text{-}R}$ subspace, but here only one 2D IRREP is allowed, which exclude any extra symmetry protected Dirac points. The whole $4N$ band structure can thus be determined by knowing the energy-ordering of the IRREPs at the high-symmetry points $\{ \Gamma, S,T,U\}$. Figure~\ref{fig_SG19_8B} gives two eight-band examples where these rules give the full band structure.

\paragraph{Four-band topology}  
Having demonstrated the existence of Dirac band-crossing points, we turn to fully characterizing their topology. We start with the simplest four-band case. For this we derive the Chern number of each Dirac point algebraically, i.e.~with  no other assumption than that the system satisfies SG19. In the following we arbitrarily split the four bands into two valence bands (occupied) and two conduction bands (unoccupied) over the whole BZ \footnote{Strictly speaking this band splitting procedure would require a $\boldsymbol{k}$-dependent Fermi level, but as a conceptual tool it is still valid.}. 

\begin{figure}[t]
\centering
\begin{tabular}{cc} 
	\includegraphics[width=0.95\linewidth]{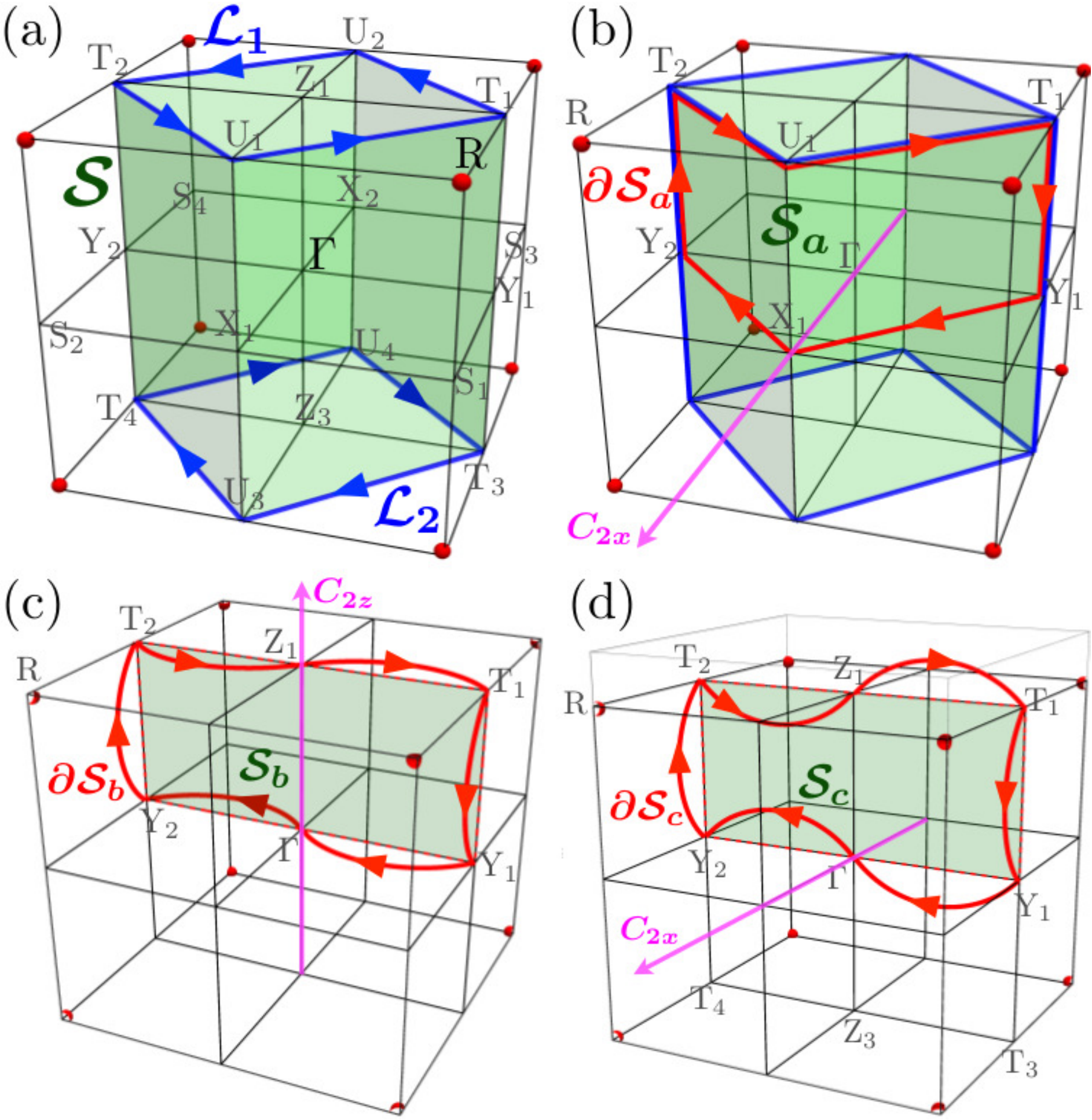}
\end{tabular} 
\caption{\label{fig_SG19_B} (a) Closed surface $\mathcal{S}$ separating subspaces $\mathcal{B}_{\Gamma}$ and $\mathcal{B}_{R}$ with the oriented boundary $\partial\mathcal{S} = \mathcal{L}_1 + \mathcal{L}_2$. (b) Oriented boundary $\partial \mathcal{S}_a$ of a subset $\mathcal{S}_a$ with $\mathcal{S} = \cup_{g\in D_2} g \mathcal{S}_a$. (c) Oriented boundary $\partial  \mathcal{S}_b$ of the closed surface $\mathcal{S}' = \mathcal{S}_b + C_{2z} \mathcal{S}_b$ surrounding $h\overline{\Gamma Z}$. (d) Oriented boundary $\partial  \mathcal{S}_c$ of the closed surface $\mathcal{S}'' =  \cup_{g\in D_2} g \mathcal{S}_c$ surrounding the plane $\overline{X}$. In (c,d) $\mathcal{S}_{(b,c)}$ is obtained from the green plane through a smooth inflation out of plane with the oriented boundary $\partial \mathcal{S}_{(b,c)}$ constrained by the symmetry requirement that $\mathcal{S}_{(b,c)} + C_{2(z,x)} \mathcal{S}_{(b,c)}$ is closed.}
\end{figure}

\begin{figure}[t]
\centering
\begin{tabular}{c} 
	\includegraphics[width=0.95\linewidth]{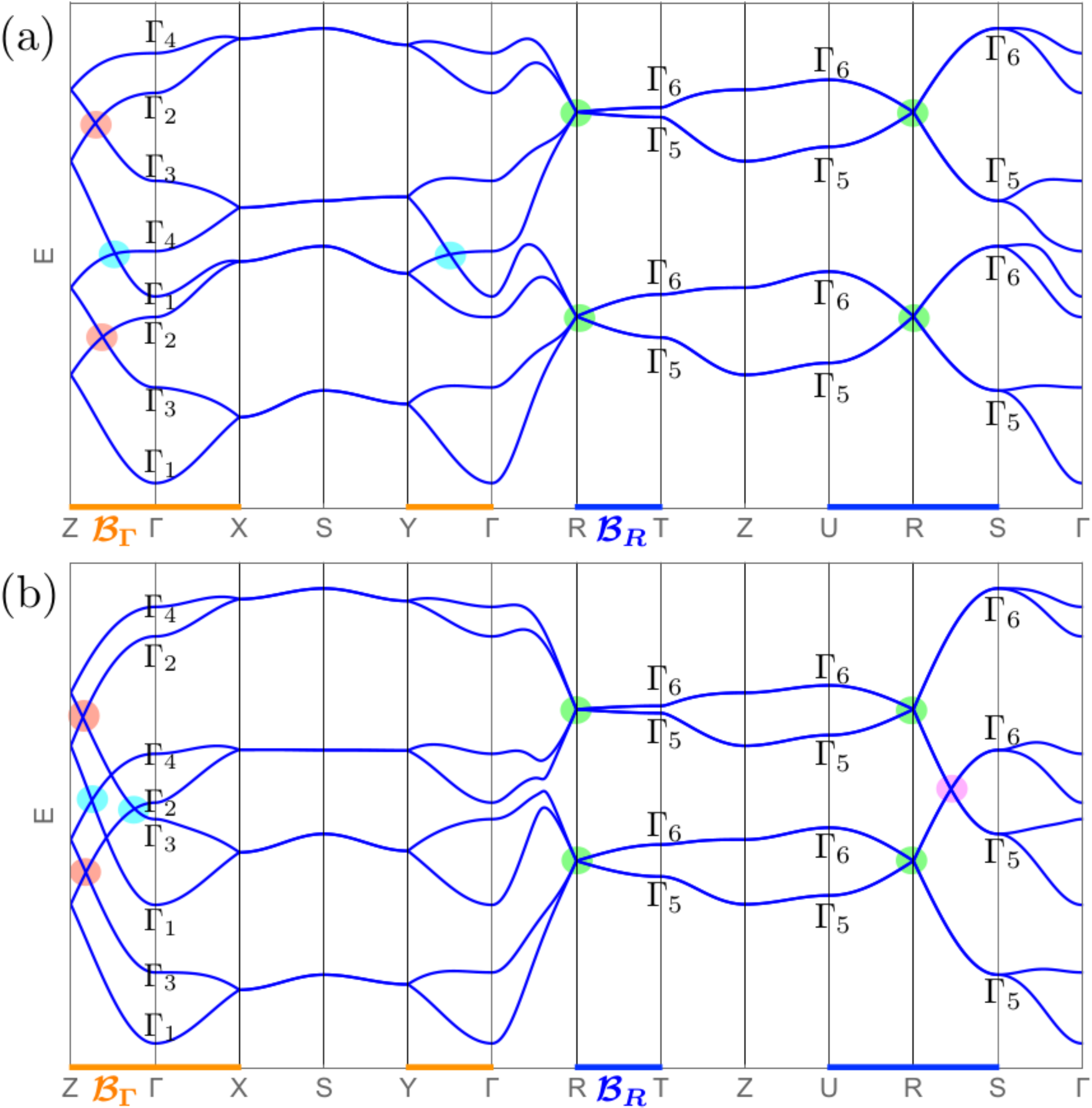} 
\end{tabular}
\caption{\label{fig_SG19_8B} Electronic band structure of eight-band tight-binding models in SG19, corresponding to (a)  class $\left( \Gamma_{II},S_{I}, T_{I} , U_{I}\right)$ and (b) class $\left( \Gamma_{III},S_{II}, T_{I} , U_{I}\right)$. Symmetry protected single Dirac points ($\vert C_1\vert=1$) globally unavoidable in red (as obtained in the four-band subspace) and class dependent in blue for an eight-band subspace, as well as double Dirac points ($\vert C_1\vert=2$) globally unavoidable in green (as in the four-band subspace) and class dependent in purple (eight-band subspace). 
}
\end{figure}

Let us first separate the two subspaces $\mathcal{B}_{\Gamma}$ and $\mathcal{B}_{R}$ by the green box $\mathcal{S}$ shown in Fig.~\ref{fig_SG19_B}(a). $\mathcal{S}$ is chosen such that it is closed (the oriented boundaries $\partial \mathcal{S}=\mathcal{L}_1 + \mathcal{L}_2 \cong 0$ due to periodicity), symmetric under $D_2$, and supporting a fully gapped spectrum. Effectively, $\mathcal{S}$ surrounds $\mathcal{B}_{\Gamma}$. Any smooth deformation of $\mathcal{S}$ satisfying these conditions and conserving the vertices also works. By Stoke's theorem the Chern number over the closed manifold $\mathcal{S}$ is simply given by the Berry phase over its boundary $\partial \mathcal{S} $, i.e.~$2\pi C_1 [\mathcal{S}] = \gamma[\partial \mathcal{S}] $. Next, we rewrite $\mathcal{S}$ as the orbit of a subset $\mathcal{S}_a$ under the point group $D_2$, i.e.~$\mathcal{S} = \bigcup_{g\in D_2} g\mathcal{S}_a$. Using the symmetry of the Berry curvature under $D_2$
\footnote{Writing the Berry curvature as $\boldsymbol{\mathcal{F}}$ and the (matrix) Berry-Wilczeck-Zee connection as $\boldsymbol{\mathcal{A}}$, we have $2\pi C_1 [\mathcal{S}] = \int_{\mathcal{S}} \boldsymbol{\mathcal{F}} \cdot d\boldsymbol{s} = \sum_{g\in D_2} \int_{g \mathcal{S}_a}  \boldsymbol{\mathcal{F}} \cdot d\boldsymbol{s} = 4 \int_{\mathcal{S}_a} \boldsymbol{\mathcal{F}} \cdot d\boldsymbol{s} = 4 \oint_{\partial \mathcal{S}_a} \mathrm{Tr}~\boldsymbol{\mathcal{A}} \cdot d\boldsymbol{l}  = 4 \gamma[\partial\mathcal{S}_a]$.}, 
we then have the simplification $2\pi C_1 [\mathcal{S}] = 4 \gamma[\partial\mathcal{S}_a]$, with $\partial\mathcal{S}_a$ the red oriented loop shown in Fig.~\ref{fig_SG19_B}(b). We are thus left with the task of evaluating the Berry phase $\gamma[\partial\mathcal{S}_a]$. Notice that we here have to assume a smooth gauge, such that the Berry phase $\gamma$ varies smoothly as we sweep the loop $\mathcal{L}$ over $\mathcal{S}$~\footnote{The smooth gauge guarantees that if $\mathcal{S}$ is a non-trivial manifold, i.e.~surrounding a topologically stable band-crossing, then $\gamma[ \mathcal{L}_1 ]$ and $\gamma[\mathcal{L}^{-1}_2]$ belong to different sectors (separated by $2n\pi$) and thus the phase difference $\gamma[ \mathcal{L}_1 ] +\gamma[ \mathcal{L}_2 ] = \gamma[ \mathcal{L}_1 ] - \gamma[ \mathcal{L}^{-1}_2 ] $ is not trivially zero, even though $\mathcal{L}_1 + \mathcal{L}_2 \cong 0$, leading to a Chern number $C_1=n$. See also [36] and [40].}.

The total abelian Berry phase of the valence (occupied) subspace of a closed loop $\mathcal{L}$ is given by $\mathrm{e}^{-i \gamma[\mathcal{L}]} = \mathrm{det}~ \mathcal{W}[\mathcal{L}]$, 
where $\mathcal{W}[\mathcal{L}]$ is the matrix (non-abelian) Wilson loop computed in the valence band basis $\vert \boldsymbol{u}_{occ},\boldsymbol{k} \rangle = \left( \vert u_{v_1},\boldsymbol{k} \rangle,\vert u_{v_2},\boldsymbol{k} \rangle\right)^T$ \cite{SM}. We can then decompose the loop $\partial\mathcal{S}_a$ into segments with high-symmetry endpoints: $\mathcal{W}[\partial\mathcal{S}_a] = \mathcal{W}_{X_1\leftarrow Y_1}  \mathcal{W}_{Y_1\leftarrow T_1}  \mathcal{W}_{T_1\leftarrow U_1}  \mathcal{W}_{U_1\leftarrow T_2}\mathcal{W}_{T_2\leftarrow Y_2}\mathcal{W}_{Y_2\leftarrow X_1}$, see Fig.~\ref{fig_SG19_B}(b). A symmetry reduction based on $C_{2x}$, assuming that the half $\mathcal{S}_a + C_{2x} \mathcal{S}_a$ of $\mathcal{S}$ is closed, using techniques developed in \cite{Bernevig1,Bernevig2,Bernevig3,Bernevig4} then gives \cite{SM}
\begin{align}
	\mathrm{det}~ \mathcal{W}[\partial\mathcal{S}_a] &= \mathrm{det}\left[\breve{S}^{X}_{x} \cdot  (\breve{S}^{Y}_{x})^{-1} \cdot \breve{S}^{T}_{x}  \cdot (\breve{S}^{U}_{x})^{-1} \right]\nonumber \\
	&= \prod\limits_{n=1,2} \dfrac{\lambda^{X}_{x,n} \lambda^{T}_{x,n}}{\lambda^{Y}_{x,n}  \lambda^{U}_{x,n} }.
	\end{align}
Here $\breve{S}^{\boldsymbol{k}^*}_{x} = \langle \boldsymbol{u}_{occ},C_{2x}\boldsymbol{k}^* \vert \{C_{2x}\vert \boldsymbol{\tau}_x\} \vert \boldsymbol{u}_{occ},\boldsymbol{k}^* \rangle$ gives a representation of the symmetry operator $\{C_{2x}\vert \boldsymbol{\tau}_x\}$ in the occupied band basis at the high-symmetry point $ C_{2x} \boldsymbol{k}^* =  \boldsymbol{k}^* - \boldsymbol{K}[C_{2x}]$ ($\boldsymbol{K}[C_{2x}]$ is a reciprocal lattice vector, possibly zero) with $\lambda^{\boldsymbol{k}^*}_{x,n}$ being its $n$th eigenvalue. Since the eigenvalues are invariant under (unitary) basis changes we can readily use tabulated IRREPs \cite{BradCrack, Bilbao} and we find $\mathrm{e}^{-i \gamma[\partial\mathcal{S}_a]}=\mathrm{det}~ \mathcal{W}[\partial\mathcal{S}_a] = -1$ and thus the Chern number
$C_1[\mathcal{S}] = 4 \gamma[\partial\mathcal{S}_a]/(2\pi) = 2\mod 4$~\footnote{The Chern number is here obtained through an absolute Berry phase, hence inheriting the torsor structure of a phase, i.e.~without a favored trivial element as the mod~4 ambiguity shows. See the SM \cite{SM} for (numerical) calculations of the Chern number through the total flow of the Berry phase, i.e.~now a phase difference, eliminating this ambiguity.}.
We thus conclude that there is a symmetry protected obstruction to the realization of a trivial insulating band structure over $\mathcal{S}$, leading to the existence of topologically stable Dirac points in $\mathcal{B}_{\Gamma}$. Choosing the smallest Berry phase, we get $C_1[\mathcal{B}_{\Gamma}] = \pm 2$. By the cancellation of the global charge (Nielsen-Ninomiya theorem \cite{Nielsen1,Nielsen2,Witten_lect}), the Chern number of $\mathcal{B}_{R} \subset BZ\backslash \mathcal{B}_{\Gamma}$ must then be $C_1[\mathcal{B}_{R}] = \mp 2$ and hence the $R$-point is necessarily a double Dirac point.

We now establish that the above result can be refined by choosing tighter ``boxes". First we consider a closed surface $\mathcal{S}'$ that surrounds half of a line of $\mathcal{B}_{\Gamma}$ (written $h\overline{\Gamma X_i}$) and is symmetric under a $C_{2i}$ rotation around that axis, see e.g.~Fig.~\ref{fig_SG19_B}(c) for $\mathcal{S}'=\mathcal{S}_b+C_{2z} \mathcal{S}_b$ surrounding $h\overline{\Gamma Z}$. Following a similar line of thought as above, the Chern number is given by the Berry phase of the surface boundary. Further proceeding with a symmetry reduction of the Wilson loop based on $C_{2i}$, we find~\cite{SM} 
\begin{eqnarray}
\label{Chern_4B_b}
	\mathrm{e}^{-i \pi C_1[h\overline{\Gamma X_i}]}  =  -  \chi^{v_1}_{i} \chi^{v_2}_{i} \;,
\end{eqnarray}
where $\chi^{v_n}_{i}$(=$\lambda^{\boldsymbol{0}}_{i,v_n}$) is the character of the 1D IRREP $\Gamma^{\boldsymbol{0}}_{v_n}(\{C_{2i}\vert \boldsymbol{\tau}_{i}\}) $ of the valence band $v_n$ at $\Gamma$, given in Fig.~\ref{fig_SG19}(b). Therefore, depending on the valence IRREPs at $\Gamma$, we either have $C_1[h\overline{\Gamma X_i}] = 0 \mod 2$ such that no Dirac point is present, or $C_1[h\overline{\Gamma X_i}] = 1 \mod 2$ which gives the existence of a simple Dirac point $(\vert C_1\vert=1)$ on the half line $h\overline{\Gamma X_i}$.

Next we instead choose a closed surface $\mathcal{S}''=\bigcup_{g\in D_2} g\mathcal{S}_c$ that surrounds the plane containing $\Gamma$ and perpendicular to the line $\overline{\Gamma X_i}$ (written $\overline{X_i}$). The Chern number is then given by $2\pi C_1[\overline{X_i}] = 4\gamma[\partial\mathcal{S}_c]$, as illustrated for $\overline{X}$ in~Fig.~\ref{fig_SG19_B}(d). Finally, the symmetry reduction of the Wilson loop based on $C_{2i}$ assuming that the half $\mathcal{S}_c+C_{2i} \mathcal{S}_c$ of $\mathcal{S}''$ is closed, gives~\cite{SM}
\begin{equation}
\label{Chern_4B_c}
	 \mathrm{e}^{-i \frac{\pi C_1\left[\overline{X}_i \right]}{2}}  = + \chi^{v_1}_{i} \chi^{v_2}_{i} \;.
\end{equation}
Thus, depending on the valence IRREPs at $\Gamma$, we either have $C_1\left[\overline{X}_i \right] = 0 \mod 4$. i.e.~no Dirac points on $\overline{X}_i$, or $C_1\left[\overline{X}_i \right] = 2 \mod 4$, demonstrating that the two simple Dirac points on the plane $\overline{X_i}$ have the same charge. This result also directly implies that the $R$ Dirac point has charge $\mp 2$. This fully characterize the global band topology of any four-band subspace with SG19~\footnote{The SM \cite{SM} contains a numerical computation of the Chern number of simple and double Dirac points for an explicit four-band tight-binding model, fully agreeing with the general results.}. 

\paragraph{Eight-band topology} 
We next consider the topology of eight bands. Similarly to before, we arbitrarily split the bands into four valence and four conduction bands over the whole BZ. We then find the Chern numbers corresponding to Eqs.~(\ref{Chern_4B_b}-\ref{Chern_4B_c}) as
\begin{align}
\label{Chern_8B_a}
	 \mathrm{e}^{-i \pi C_1 \left[h\overline{\Gamma X_i}\right] } & =  
	  (-1)^2 \chi^{v_1}_{i} \chi^{v_2}_{i} \chi^{v_3}_{i} \chi^{v_4}_{i}, \\
	  \label{Chern_8B_b}
 \mathrm{e}^{-i  \pi \frac{ C_1 \left[ \overline{X_i} \right] }{2}} & = 
	 (+1)^2 \chi^{v_1}_{i} \chi^{v_2}_{i} \chi^{v_3}_{i} \chi^{v_4}_{i}.
\end{align}
Both are thus still determined by the valence IRREPs at $\Gamma$. From Eqs.~(\ref{Chern_8B_a}-\ref{Chern_8B_b}) we identify three topologically inequivalent classes of band structures over $\mathcal{B}_{\Gamma}$. This leads to Table \ref{eight_SG19} \footnote{$\Gamma_{III}$ is found by taking the square-root of Eqs.~(\ref{Chern_8B_a}-\ref{Chern_8B_b}).}. $\Gamma_I$ excludes simple Dirac points within $\mathcal{B}_{\Gamma}$. $\Gamma_{II}$ enforces two pairs of same-charge simple Dirac points within $\mathcal{B}_{\Gamma}$ with a total charge of $0$ or $\vert 4\vert$. $\Gamma_{III}$ has one quadruple of (same-charge) simple Dirac points on a single line in $\mathcal{B}_{\Gamma}$. 
{\def\arraystretch{1.3}  
\begin{table}[ht]
 \begin{tabular}{|l|c|} 
\hline 

$\Gamma_I$:$\left\{\Gamma_1,\Gamma_2,\Gamma_3,\Gamma_4\right\}$ & $C_1 \left[h \overline{\Gamma X_i} \right] = \frac{C_1 \left[ \overline{X_i} \right] }{2} =0\mod 2$  \\
 \hline
$\Gamma_{II}$:$\left\{ \Gamma_j,\Gamma_j, \Gamma_k,\Gamma_{l}  \right\}_{j\neq k \neq l}$ & $\mathrm{e}^{-i  \pi  C_1 \left[ h\overline{\Gamma X_i} \right] } = \mathrm{e}^{-i  \pi \frac{ C_1 \left[ \overline{X_i} \right] }{2} }=\chi^{k}_{i}\chi^{l}_{i}$ \\
 \hline
$\Gamma_{III}$:$\left\{\Gamma_j,\Gamma_j,\Gamma_k,\Gamma_k \right\}_{j\neq k}$  &  
		$\begin{array}{c} \mathrm{e}^{-i  \pi \frac{ C_1 \left[ h\overline{\Gamma X_i} \right] }{2} } = -\chi^{j}_{i}\chi^{k}_{i} \\
		\mathrm{e}^{-i  \pi \frac{ C_1 \left[ \overline{X_i} \right] }{4} } = +\chi^{j}_{i}\chi^{k}_{i}
			\end{array} $\\
\hline
 \end{tabular}
\caption{\label{eight_SG19} The three inequivalent band topologies of $\mathcal{B}_{\Gamma}$ given by the Chern numbers of its half-lines ($h\overline{\Gamma X}_i$) and planes ($\overline{X}_i$) for eight bands.}
\end{table} 
}

We also have to characterize the topology of $\mathcal{B}_{R}$. We find for the three half-lines in $\mathcal{B}_{R}$ (written $\{h\overline{RU_i}\}$) \cite{SM}:
\begin{equation}
\label{8B_Chern_c}
	\mathrm{e}^{-i \pi C_1 \left[h\overline{R U_i}\right] } =  
	 (-1)^{2} \mathrm{det}\left[ \Gamma^{U_i}_{v_1}\left(C_{2i}\right) \Gamma^{U_i}_{v_2}\left(C_{2i}\right) \right]\;.
\end{equation}
Thus two inequivalent situations can be realized according to if $\Gamma^{U_i}_{v_1} = \Gamma^{U_i}_{v_2}$ or $\Gamma^{U_i}_{v_1} \neq \Gamma^{U_i}_{v_2}$. From this we derive Table \ref{eight_SG19_R} \footnote{$U_{i,II}$ is found by taking the square-root of Eq.~(\ref{8B_Chern_c}).}. In both cases the Chern numbers of the planes containing $R$ and perpendicular to the axes $\{\overline{RU_i}\}$ (written $\overline{U_i}$) are \cite{SM}
%
$	C_1 \left[\overline{U_i}\right] = 0 \mod 4$.
%
Thus, the class $U_{i,I}$ corresponds to a band structure with no double Dirac point on the half-line $h\overline{RU_i}$, while the class $U_{i,II}$ has a pair of (same-charge) double Dirac points on the line $\overline{RU_i}$.
{\def\arraystretch{1.4} 
\begin{table}[ht]
 \begin{tabular}{|l|c|} 
\hline 
$U_{i,I}$:~$\left\{\Gamma_5,\Gamma_{6}\right\}$ & $C_1 \left[ h\overline{R U_i} \right] = \frac{C_1 \left[ \overline{U_i} \right] }{2} =0\mod 2$  \\
 \hline
$U_{i,II}$:~$\left\{\Gamma_j,\Gamma_j\right\}$
& $C_1 \left[ h\overline{R U_i} \right] = \frac{C_1 \left[ \overline{U_i} \right] }{2} =2\mod 4$ \\
\hline 
 \end{tabular}
\caption{\label{eight_SG19_R} The two inequivalent band topologies of $\mathcal{B}_{R}$ given by the Chern numbers of its half-lines ($h\overline{RU}_i$) and planes ($\overline{U}_i$) for eight bands.}
\end{table}
} 

The band structure of any eight bands is fully characterized in terms of the classes $\{\Gamma_{I},\Gamma_{II},\Gamma_{III}\}$ for $\mathcal{B}_{\Gamma}$ and $\{S_{I},S_{II},T_{I},T_{II},U_{I},U_{II}\}$ for $\mathcal{B}_{R}$.
The classes are in turn determined by the energy ordered IRREPs at the high-symmetry points $\{\Gamma,S,T,U\}$. 
While we argued that the two subspaces $\mathcal{B}_{\Gamma}$ and $\mathcal{B}_{R}$ are symmetry independent, they are actually constrained by a global charge cancellation over the whole BZ. Enumerating all the combinatorial possibilities up to the charge cancellation requirement results in the eight inequivalent band topological classes presented in Table~\ref{TOP_classes_8B}.
\begin{table}[t]
 \begin{tabular}{|c|c|c|} 
\hline 
	$\Gamma_{I}$ & $\Gamma_{II}$ & $\Gamma_{III}$ \\
 \hline
	$(S_{I},T_{I},U_{I})$ &~ $(S_{I},T_{I},U_{I})$ ~&~ $\left[(S_{II},T_{I},U_{I})\right]~$ \\
	~$\left[(S_{II},T_{II},U_{I})\right]$~ & $\left[(S_{II},T_{I},U_{I})\right]$ & $(S_{II},T_{II},U_{II})$ \\
	  & ~$\left[(S_{II},T_{II},U_{I})\right]$~ &   \\
	  & $(S_{II},T_{II},U_{II})$ &   \\
 \hline
 \end{tabular}
\caption{\label{TOP_classes_8B} The eight nonequivalent band topological classes for eight bands. $\left[ (\cdot) \right]$ means the equivalence class obtained by cyclic permutations of $S,T,U$.
}
\end{table}

A band structure example from class $(\Gamma_{II},S_{I}, T_{I} , U_{I})$ is shown in Fig.~\ref{fig_SG19_8B}(a). It realizes two pairs of simple Dirac points (blue); one pair on $\overline{\Gamma Y}$ (both with charge $\pm 1$) and one on $\overline{\Gamma Z}$ (both with charge $\mp 1$). These new Dirac points are in addition to those found above within the four band subspaces (red and green). 
Density functional theory (DFT) band structures of 3D organic materials with SG19 belonging to the same class has also recently been found~\cite{GelBoBoBa_II}. 
Another example from class $(\Gamma_{III},S_{II}, T_{I} , U_{I})$ is shown in Fig.~\ref{fig_SG19_8B}(b). It realizes a quadruple of simple Dirac points (blue) on $\overline{\Gamma Z}$ (all of charge $\mp 1$) and a pair of double Dirac points (purple) on $\overline{RS}$ (both of charge $\pm 2$).

The technique illustrated above can be straightforwardly extended to arbitrary $4N$-band structures. Simply from the list of energy ordered IRREPs at $\{\Gamma,S,T,U\}$ we can thus deduce the global band topology and predict all simple and double Dirac points protected by symmetry. 
In addition, at integer filling the number of valence bands must be a multiple of two and the IRREP ordering determines if the material is an insulator (the filling number is then a multiple of four) or a topological (semi-)metal, where the Fermi level necessarily crosses the bands forming Dirac points. 
In fact, due to global charge cancellation, distinct sets of Dirac points, not connected by point symmetry, are necessarily realized, such that the Fermi level generally does not cross at the point nodes, see e.g.~Fig.~\ref{fig_SG19}(c). This leads to ``Dirac-point metals" (the point-nodal analogue to \cite{Thomas_line}). A true semi-metallic state is only achieved through fine tuning of the band structure.
The generality of the technique also means that it is not restricted to SG19 and it can also easily be generalized to include spin degrees of freedom. 
The strong explicative and predictive power in combining symmetry and topology makes this algebraic approach
highly complementary to current data mining searches for topological semimetals \cite{BorysovGeilhufeBalatsky_mining,Vishwanath_mining} and can place it at the core of future intelligent DFT-based data mining schemes.


\begin{center}
\textbf{ACKNOWLEDGEMENTS}
\end{center}
We thank A.~Balatsky, S.~Borysov and M.~Geilhufe for useful discussions. The first author thank A. Furusaki for an insightful discussion in the early stage of the work. This work was supported by the Swedish Research Council (Vetenskapsr\aa det) and the Knut and Alice Wallenberg Foundation.

\paragraph{Note added} During the completion of this work, the preprints \cite{1612.02007,1701.08725} considering global topological classification also appeared. While some conclusions are similar to ours (especially the former that also considers a 3D example) their derivations are more formal (making use of K-theory) and they both exclude time reversal symmetry. Our approach is more heuristic and importantly includes time-reversal symmetry, which is usually present in the normal state.


\clearpage

\begin{widetext}

\begin{center}
\textbf{SUPPLEMENTARY MATERIAL}
\end{center}

\tableofcontents

\section{Symmetry based model with SG19}

In this section we introduce the symmetry based construction of tight-binding models for the space group \#19 (SG19). We review the representation of space group symmetry transformations in  second quantization and summarize the irreducible representations (IRREPs) at the high-symmetry points and lines of the Brillouin zone (BZ). This section sets the notations and gives the definitions of the standard tools of space group representation theory and tight-binding modeling that is used in the main text.

\subsection{Symmetry decomposition of the degrees of freedom}\label{4N_dof}
A possible presentation of the space group $\mathcal{G}$ = P$2_12_12_1$ (SG19) is given through the left coset with respect to its normal subgroup of translations $\mathcal{T}$ (the primitive orthorhombic Bravais lattice):
\begin{equation}
\label{left_coset}
	P2_12_12_1 = 
	\{E\vert \boldsymbol{0}\} \mathcal{T} \oplus \{C_{2z}\vert \boldsymbol{\tau}_z \} \mathcal{T}
	\oplus \{C_{2y}\vert \boldsymbol{\tau}_y \} \mathcal{T}
	\oplus \{C_{2x}\vert \boldsymbol{\tau}_x \} \mathcal{T}\;,  
\end{equation}
where the three screw symmetries $\{g\vert \boldsymbol{\tau}_g\}$ are formed with the point group elements $g\in D_2 = \{E,C_{2z},C_{2y},C_{2x}\}$ and the fractional translations $ \boldsymbol{\tau}_x = (\boldsymbol{a}_1+\boldsymbol{a}_2)/2$, $ \boldsymbol{\tau}_y = (\boldsymbol{a}_2+\boldsymbol{a}_3)/2$ and $ \boldsymbol{\tau}_z = (\boldsymbol{a}_1+\boldsymbol{a}_3)/2$ given in terms of the primitive basis $\{\boldsymbol{a}_1,\boldsymbol{a}_2,\boldsymbol{a}_3\}$. The coset representatives of Eq.~(\ref{left_coset}) have been chosen such that they are all defined with respect to a unique origin. 
We take $\boldsymbol{a}_1 = (a,0,0)$, $\boldsymbol{a}_2 = (0,b,0)$, and $\boldsymbol{a}_3 = (0,0,c)$, with the primitive reciprocal lattice vectors, 
\begin{equation}
\left\{
\begin{aligned}
	\boldsymbol{b}_1 &= 2\pi \dfrac{\boldsymbol{a}_2\times \boldsymbol{a}_3 }{(\boldsymbol{a}_1\cdot \boldsymbol{a}_2 \times \boldsymbol{a}_3)}\;, \\
	\boldsymbol{b}_2 &= 2\pi \dfrac{\boldsymbol{a}_3\times \boldsymbol{a}_1 }{(\boldsymbol{a}_1\cdot \boldsymbol{a}_2 \times \boldsymbol{a}_3)}\;, \\
	\boldsymbol{b}_3 &= 2\pi \dfrac{\boldsymbol{a}_1\times \boldsymbol{a}_2 }{(\boldsymbol{a}_1\cdot \boldsymbol{a}_2 \times \boldsymbol{a}_3)}\;.
\end{aligned}\right.
\end{equation} 

SG19 has only one Wyckoff position with no symmetry leading to an intrinsic fourfold dimensionality, i.e.~the order of the orbit of one site under the action of the four coset representatives in Eq.~(\ref{left_coset}). Therefore, assuming $4N$ sub-lattice sites within the primitive cell, the decomposition of the sub-lattice space $\Gamma^{sl}$ into IRREPs of $D_2$ (see the character table in Fig.~1(b) in the main text) is generically given by $\Gamma^{4N} = N\left(\Gamma_1 \oplus \Gamma_2 \oplus \Gamma_3 \oplus \Gamma_4\right) $. Also, the space of electronic orbitals decomposes as $\Gamma^{e.o.}=\bigoplus\limits_{l=0,1,2,\dots} \Gamma^{l} = \bigoplus\limits_{l,j} n^l_j \Gamma_j $ where the set of integers $\{n^l_j\}_{j=1,2,3,4}$ characterizes the decomposition of the $l$-th electronic orbital into IRREPs of $D_2$. We thus conclude that any basis for SG19 (neglecting spin-orbit coupling (SOC)) splits as 
\begin{equation}
\label{sym_decomp}
	\Gamma^{e.o.} \times \Gamma^{4N} = 
	\bigoplus\limits_{l,j} n^l_j  \left(\Gamma_1 \oplus \Gamma_2 \oplus \Gamma_3 \oplus \Gamma_4 \right) \;.
\end{equation}

\subsection{Tight-binding basis and construction of the Hamiltonian}

We here build explicitly the four tight-binding basis functions of a four-dimensional (4D) subspace$-$let us call it $(1)$$-$, i.e. $\Gamma^{(1)}_{1} \oplus \Gamma^{(1)}_{2} \oplus \Gamma^{(1)}_{3} \oplus \Gamma^{(1)}_{4}$, obtained after the symmetry decomposition of Eq.~(\ref{sym_decomp}) is performed over the total $4N$ dimensional space. The four basis functions must correspond to the four sub-lattice sites belonging to the same Wyckoff's position such that they are mapped onto one another under the left-coset representatives of Eq.~(\ref{left_coset}). 

From the Wannier functions at the $m$-th unit cell $\boldsymbol{R}_m \in \mathcal{T}$, $\vert w_i , \boldsymbol{R}_m+\boldsymbol{r}_i \rangle$ for $i=1,2,3,4$ (with $\boldsymbol{r}_1,\boldsymbol{r}_2,\boldsymbol{r}_3,\boldsymbol{r}_4$, the positions of four representative sites), we first form the Bloch-L\"owdin basis (or site basis) of this subspace
\begin{equation}
\label{site_basis}
	\vert \boldsymbol{\varphi} , 1  , \boldsymbol{k} \rangle = \left(\begin{array}{c}  \vert \phi_{\boldsymbol{r}_1}  , \boldsymbol{k} \rangle \\
	 \vert \phi_{\boldsymbol{r}_2}  , \boldsymbol{k} \rangle \\
	  \vert \phi_{\boldsymbol{r}_3}  , \boldsymbol{k} \rangle \\
	   \vert \phi_{\boldsymbol{r}_4} , \boldsymbol{k} \rangle \end{array}\right)^T = \dfrac{1}{\sqrt{N}}\sum\limits_{\boldsymbol{R}_m} \left(\begin{array}{c} \mathrm{e}^{i \boldsymbol{k}\cdot (\boldsymbol{R}_m+\boldsymbol{r}_{1})}	\vert w_1 , \boldsymbol{R}_m+\boldsymbol{r}_{1} \rangle \\
	\mathrm{e}^{i \boldsymbol{k}\cdot (\boldsymbol{R}_m+\boldsymbol{r}_{2})}	\vert w_2, \boldsymbol{R}_m+\boldsymbol{r}_{2} \rangle \\
	\mathrm{e}^{i \boldsymbol{k}\cdot (\boldsymbol{R}_m+\boldsymbol{r}_{3})}	\vert w_3, \boldsymbol{R}_m+\boldsymbol{r}_{3} \rangle \\
	\mathrm{e}^{i \boldsymbol{k}\cdot (\boldsymbol{R}_m+\boldsymbol{r}_{4})}	\vert w_4, \boldsymbol{R}_m+\boldsymbol{r}_{4} \rangle
	\end{array}\right)^T\;,
\end{equation}
($N$ is the total number of those selected sites that belong to the same Wyckoff's position). Then we form the symmetrized basis, we call it the symmetry-Bloch-L\"owdin basis,
\begin{eqnarray}
\label{sym_basis}
	\vert \boldsymbol{\phi},1 ,\vk \rangle &=& \left(\begin{array}{c}   \vert \phi_{\Gamma_1},1 ,\vk \rangle \\
	 \vert \phi_{\Gamma_2},1 ,\vk \rangle \\
	  \vert \phi_{\Gamma_3} ,1,\vk \rangle \\
	   \vert \phi_{\Gamma_4},1 ,\vk \rangle \end{array}\right)^T 
	   = \vert \boldsymbol{\varphi} , 1 , \boldsymbol{k} \rangle \cdot \hat{U}^S \;,
\end{eqnarray}
where then 
\begin{equation}
	\hat{U}^S = \dfrac{1}{2}  \left(\begin{array}{rrrr} 1 & 1 & 1 & 1\\
	1& -1 & -1 & 1 \\ 
	1 & 1 & -1 & -1 \\
	1 & -1 & 1 & -1
	\end{array}\right) \;.
\end{equation}
The symmetry-Bloch-L\"owdin basis functions of all the other 4D subspaces of Eq.~(\ref{sym_decomp}) can be built similarly, giving the total symmetry-Bloch-L\"owdin basis, 
\begin{equation}
	\vert \boldsymbol{\phi} ,\vk \rangle = \left(\begin{array}{c} \vert  \boldsymbol{\phi} , 1 ,\boldsymbol{k} \rangle \\
		 	\vdots\\
			\vert  \boldsymbol{\phi} , N ,\boldsymbol{k} \rangle 
	\end{array}\right)^T	\;,
\end{equation}
for a system of $4N$ degrees of freedom

The total Hamiltonian for a system of $4N$ degrees of freedom then takes the following $(4\times 4)$-block structure in the symmetry-Bloch-L\"owdin basis,
\begin{equation}
 \mathcal{H} = \sum_{\boldsymbol{k} \in \mathrm{BZ}} \sum\limits_{n,m =1}^{N}  \vert \boldsymbol{\phi},n ,\vk \rangle H_{nm}(\vk) \langle \boldsymbol{\phi},m ,\vk \vert \;,
\end{equation}
such that a $(4\times 4)$-block Hamiltonian has the structure   
\begin{equation}
 	H_{nm}(\vk) = \left[ \begin{array}{cccc}	h^{(\Gamma_1)}_{\Gamma_1\Gamma_1}(\vk) & h^{(\Gamma_2)}_{\Gamma_1\Gamma_2}(\vk) & h^{(\Gamma_3)}_{\Gamma_1\Gamma_3}(\vk) & h^{(\Gamma_4)}_{\Gamma_1\Gamma_4}(\vk)\\
		 & h^{(\Gamma_1)}_{\Gamma_2\Gamma_2}(\vk) & h^{(\Gamma_4)}_{\Gamma_2\Gamma_3}(\vk) & h^{(\Gamma_3)}_{\Gamma_2\Gamma_4}(\vk)\\
		 	& & h^{(\Gamma_1)}_{\Gamma_3\Gamma_3}(\vk) & h^{(\Gamma_2)}_{\Gamma_3\Gamma_4}(\vk)\\
			& & & h^{(\Gamma_1)}_{\Gamma_4\Gamma_4}(\vk)\\
	\end{array}\right]_{nm} \;,
\end{equation}
where the remaining off-diagonal part is given through hermiticity. Here some explanations are needed: $[h^{(\Gamma_j)}_{\Gamma_a\Gamma_b}(\vk)]_{nm}$ is the matrix element of the Hamiltonian that connects the operators $\vert \phi_{\Gamma_a}, n, \boldsymbol{k} \rangle \langle \phi_{\Gamma_b}, m, \boldsymbol{k} \vert$, such that $h^{(\Gamma_j)}(\vk)$ is a basis function of the IRREP $\Gamma_j= \Gamma_a \times \Gamma_b $ as a function of $\boldsymbol{k}$.

We finally define the band basis
\begin{eqnarray}
\label{band_basis}
	\vert \boldsymbol{\psi} ,\boldsymbol{k} \rangle &=& \left(\begin{array}{c} \vert \boldsymbol{\psi},1 ,\boldsymbol{k} \rangle \\
		 	\vdots\\
			\vert \boldsymbol{\psi},N ,\boldsymbol{k} \rangle 
	\end{array}\right)^T = \vert \boldsymbol{\phi} ,\boldsymbol{k} \rangle \cdot \breve{U}(\boldsymbol{k}) \;,
\end{eqnarray}
which can be decomposed as
\begin{eqnarray}
	\vert \boldsymbol{\psi},n ,\boldsymbol{k} \rangle &=& \left(\begin{array}{c} \vert \psi_{\Gamma_1},n ,\boldsymbol{k} \rangle \\
		 	\vert \psi_{\Gamma_2},n ,\boldsymbol{k} \rangle \\
			 \vert \psi_{\Gamma_3},n ,\boldsymbol{k} \rangle \\
			\vert \psi_{\Gamma_4},n ,\boldsymbol{k} \rangle 
	\end{array}\right)^T  \;,\quad n=1,\dots,N\;,
\end{eqnarray}
 such that it diagonalizes the Hamiltonian, i.e.
\begin{eqnarray}
 	\mathcal{H}_{nm} &=& \sum_{\boldsymbol{k} \in \mathrm{BZ}}   \vert \boldsymbol{\psi} ,\vk \rangle \cdot \left[\breve{U}^{\dagger}(\boldsymbol{k}) \cdot H(\vk) \cdot \breve{U}(\boldsymbol{k})\right] \cdot \langle \boldsymbol{\psi} ,\vk \vert \;,\nonumber\\
		&=& \sum_{\boldsymbol{k} \in \mathrm{BZ}}   \vert \boldsymbol{\psi} ,\vk \rangle \cdot \mathrm{diag}\left( E_{\Gamma_1}^1(\boldsymbol{k}) , \cdots, E_{\Gamma_4}^N(\boldsymbol{k}) \right) \cdot \langle \boldsymbol{\psi} ,\vk \vert \;,
\end{eqnarray}
where $E_{\Gamma_j}^n$ is an eigenvalue labeled by its IRREP at $\Gamma$ with $j=1,\dots,4$ and its band number $n=1,\dots,N$.

\subsection{Symmetry transformations and space group representations}
\subsubsection{Transformation of the symmetry-Bloch-L\"owdin basis}
From Eq.~(\ref{site_basis}) and Eq.~(\ref{sym_basis}) we derive how a nonsymmorphic symmetry transformation acts on the symmetry-Bloch-L\"owdin basis of the 4D subspace $n$, see Supplementary Information of \cite{Bernevig3}, 
\begin{equation}
\label{psi_sym_a}
	^{\{g\vert \boldsymbol{\tau}_g\}}   \vert \boldsymbol{\phi},n ,\vk \rangle = \mathrm{e}^{-i g\boldsymbol{k}\cdot \boldsymbol{\tau}_g} \vert \boldsymbol{\phi},n , g\vk \rangle \cdot  \hat{U}^{(n)}_{\{g\vert \boldsymbol{\tau}_g\}}\;,
\end{equation}
with
\begin{equation}
	 \hat{U}^{(n)}_{\{g\vert \boldsymbol{\tau}_g\}} = \bigoplus\limits_{j=1,2,3,4} \chi^{\Gamma_j}(g)\;,
\end{equation}
where $\chi^{\Gamma_j}(g)$ are the characters for $D_2$ given in Fig.~1(b) of the main text. We find for a translation transformation ($\boldsymbol{t} \in \mathcal{T}$):
\begin{equation}
	^{\{E\vert \boldsymbol{t}\}} \vert \boldsymbol{\phi} ,\vk \rangle =  \mathrm{e}^{-i \boldsymbol{k}\cdot \boldsymbol{t}} \vert \boldsymbol{\phi} , \vk \rangle	\;.
\end{equation}
We then get the unitary representation of the symmetry operators $\{g\vert \boldsymbol{\tau}_g+\boldsymbol{t}\}$ in this basis,
\begin{eqnarray}
		\hat{S}^{\boldsymbol{k}}\left(\{g\vert \boldsymbol{\tau}_g+\boldsymbol{t}\}\right) & \equiv & \langle \boldsymbol{\phi} ,g\vk \vert  ^{\{g\vert \boldsymbol{\tau}_g+\boldsymbol{t}\}} \vert \boldsymbol{\phi} ,\vk \rangle \;,\nonumber\\
\label{sym_basis_trans}
		&=& \mathrm{e}^{-i g\boldsymbol{k}\cdot (\boldsymbol{\tau}_g + \boldsymbol{t}) }   \hat{U}_{\{g\vert \boldsymbol{\tau}_g\}}\;,
\end{eqnarray}
with
\begin{equation}
	\hat{U}_{\{g\vert \boldsymbol{\tau}_g\}} = \bigoplus\limits_{n=1}^{N}\hat{U}^{(n)}_{\{g\vert \boldsymbol{\tau}_g\}}	\;.
\end{equation}

Note that because of unitarity 
\begin{equation}
	\langle \boldsymbol{\phi} ,g\vk \vert \left(^{\{g\vert \boldsymbol{\tau}_g+\boldsymbol{t}\}} \vert \boldsymbol{\phi} ,\vk \rangle \right)  = \left(^{\{g\vert \boldsymbol{\tau}_g+\boldsymbol{t}\}^{\dagger}} \langle \boldsymbol{\phi} ,g\vk   \vert\right) \vert \boldsymbol{\phi} ,\vk \rangle \;.
\end{equation}
Also, in general, the representation in the symmetry-Bloch-L\"owdin basis, $\hat{S}^{\boldsymbol{k}}\left(\{g\vert \boldsymbol{\tau}_g+\boldsymbol{t}\}\right)$, is reducible.

As we will see, the transformation under reciprocal lattice translations, $\boldsymbol{K} = p \boldsymbol{b}_1 + q \boldsymbol{b}_2+ r\boldsymbol{b}_3 $, plays a non-trivial role in nonsymmorphic space groups. We have
\begin{equation}
	\vert \boldsymbol{\phi},n ,\vk+\boldsymbol{K} \rangle = \vert \boldsymbol{\phi},n ,\vk\rangle \cdot\hat{T}^{(n)}(\boldsymbol{K})	\;,
\end{equation}
with
\begin{equation}
	\hat{T}^{(n)}(\boldsymbol{K}) = \hat{U}^{S\dagger}\cdot \mathrm{Diag}\left[\mathrm{e}^{i \boldsymbol{K}\cdot \boldsymbol{r}_1},\mathrm{e}^{i \boldsymbol{K}\cdot \boldsymbol{r}_2},\mathrm{e}^{i \boldsymbol{K}\cdot \boldsymbol{r}_3},\mathrm{e}^{i \boldsymbol{K}\cdot \boldsymbol{r}_4} \right] \cdot \hat{U}^S	\;,
\end{equation}
and for the full $4N$ space we then trivially have
\begin{equation}
	\vert \boldsymbol{\phi} ,\vk+\boldsymbol{K} \rangle = \vert \boldsymbol{\phi} ,\vk\rangle \cdot \hat{T}(\boldsymbol{K})	\;,
\end{equation}
with
\begin{equation}
	\hat{T}(\boldsymbol{K}) = \bigoplus\limits_{n=1}^{N} \hat{T}^{(n)}(\boldsymbol{K}) \;.
\end{equation}
\subsubsection{Transformation of the band basis}

We now consider the action of a symmetry transformation $\{g\vert \boldsymbol{\tau}_g+\boldsymbol{t}\}$ on the band basis, which we derive from Eqs.~(\ref{band_basis}), (\ref{psi_sym_a}) and (\ref{sym_basis_trans}),
\begin{equation}
	^{\{g\vert \boldsymbol{\tau}_g+\boldsymbol{t}\}}   \vert \boldsymbol{\psi} ,\vk \rangle = \mathrm{e}^{-i g\boldsymbol{k}\cdot (\boldsymbol{\tau}_g+\boldsymbol{t})} \vert \boldsymbol{\psi} , g\vk \rangle \cdot \left[ \breve{U}^{\dagger}(g\boldsymbol{k}) \hat{U}_{\{g\vert \boldsymbol{\tau}_g\}} \breve{U}(\boldsymbol{k}) \right] \;,
\end{equation}
from which we derive a representation of the symmetry operator,
\begin{eqnarray}
	\breve{S}^{\boldsymbol{k}}\left(\{g\vert \boldsymbol{\tau}_g+\boldsymbol{t}\}\right)  &\equiv& \langle \boldsymbol{\psi} ,g\vk \vert ^{\{g\vert \boldsymbol{\tau}_g+\boldsymbol{t}\}}   \vert \boldsymbol{\psi} ,\vk \rangle  \;, \nonumber\\
\label{band_basis_trans}
	&=& \mathrm{e}^{-i g\boldsymbol{k}\cdot (\boldsymbol{\tau}_g+\boldsymbol{t})}  \cdot \left[ \breve{U}^{\dagger}(g\boldsymbol{k}) \hat{U}_{\{g\vert \boldsymbol{\tau}_g\}} \breve{U}(\boldsymbol{k}) \right]\;. 
\end{eqnarray}

At the high-symmetry points $g\boldsymbol{k}^* = \boldsymbol{k}^* - \boldsymbol{K}_g$, where $\boldsymbol{K}_g$ is a reciprocal lattice vector (possibly zero), the band representation naturally splits into the IRREPs $-$also called \textit{small representations}$-$of the little group $\mathcal{G}^{ \boldsymbol{k}^*}$,
\begin{equation}
	\breve{S}^{\boldsymbol{k}^*}\left(\{g\vert \boldsymbol{\tau}_g\}\right)  = \bigoplus\limits_{\alpha} \Gamma^{\boldsymbol{k}^*}_{\alpha}\left(\{g\vert \boldsymbol{\tau}_g\}\right) \;,
\end{equation}
which takes the generic form
\begin{equation}
	\Gamma^{\boldsymbol{k}^*}_{\alpha}\left(\{g\vert \boldsymbol{\tau}_g\}\right) = \mathrm{e}^{-i \boldsymbol{k}^*\cdot \boldsymbol{\tau}_{g}} D^{\boldsymbol{k}^*}_{\alpha}\left(\{g\vert \boldsymbol{\tau}_g\}\right)	\;,
\end{equation}
where $D^{\boldsymbol{k}^*}_{\alpha}\left(\{g\vert \boldsymbol{\tau}_g\}\right)$ can be found either as an IRREP of the little co-group $\bar{\mathcal{G}}^{ \boldsymbol{k}^*}$ when $g\boldsymbol{k}^* = \boldsymbol{k}^*$, or as projective IRREP of $\bar{\mathcal{G}}^{ \boldsymbol{k}^*}$ when $\boldsymbol{k}^*$ is on the BZ boundary such that $g\boldsymbol{k}^* = \boldsymbol{k}^* - \boldsymbol{K}_g$ with $\boldsymbol{K}_g \neq 0$, see Ref.~\cite{BradCrack} for a standard textbook on the subject.

A translation by a reciprocal vector $\boldsymbol{K}$ gives
\begin{equation}
\label{band_basis_K}
\begin{aligned}
	\vert \boldsymbol{\psi} ,\boldsymbol{k}+\boldsymbol{K} \rangle &= \vert \boldsymbol{\psi} ,\boldsymbol{k} \rangle \cdot \breve{M}_{\boldsymbol{k}}(\boldsymbol{K}) \;,\\
 	\breve{M}_{\boldsymbol{k}}(\boldsymbol{K}) &= \breve{U}^{\dagger}(\boldsymbol{k}) \cdot \hat{T}(\boldsymbol{K}) \cdot \breve{U}(\boldsymbol{k}+\boldsymbol{K}) \;.
\end{aligned}
\end{equation}
\subsubsection{Time reversal operation}

Time reversal symmetry (TRS) is given as%
\begin{equation}
		^{\Theta} \vert \boldsymbol{\phi} ,\vk \rangle = \vert \boldsymbol{\phi} , -\vk \rangle \cdot \hat{U}^{\Theta}  \mathcal{K}\;,
\end{equation}
where $\mathcal{K}$ is the complex conjugation and $\hat{U}^{\Theta}$ captures the effect of time reversal on the electronic orbitals, which can be readily deduced from its action on spherical harmonics, i.e. $^{\Theta}\vert Y_l^m\rangle = (-1)^{l-m} \vert Y_l^{-m}\rangle $ (see for instance Ref.~\cite{BrinkSatchler}).  
Therefore, in the band basis time reversal acts as
\begin{equation}
	^{\Theta} \vert \boldsymbol{\psi} ,\vk \rangle	= \vert \boldsymbol{\psi} ,-\vk \rangle \cdot \left[ \breve{U}^{\dagger}(-\boldsymbol{k})\cdot \hat{U}^{\Theta} \cdot \breve{U}^*(\boldsymbol{k}) \right] \mathcal{K}	\;,
\end{equation}
which gives a unitary representation of TRS
\begin{equation}
	\breve{\Theta}^{\boldsymbol{k}}   \equiv \langle \boldsymbol{\psi} ,-\vk \vert  ^{\Theta} \vert \boldsymbol{\psi} ,\vk \rangle =  \breve{U}^{\dagger}(-\boldsymbol{k})\cdot \hat{U}^{\Theta} \cdot \breve{U}^*(\boldsymbol{k}) 	\;,
\end{equation}
with $\breve{\Theta}^{\boldsymbol{k}}\cdot \left.\breve{\Theta}^{\boldsymbol{k}}\right.^{\dagger} = \breve{1}$. We then have
\begin{equation}
	^{\Theta^2} \vert \boldsymbol{\psi} ,\vk \rangle = \vert \boldsymbol{\psi} ,\vk \rangle \cdot \left[\breve{U}^{\dagger}(\boldsymbol{k}) \cdot  \hat{U}^{\Theta} \cdot  (\hat{U}^{\Theta})^* \cdot \breve{U}(\boldsymbol{k}) \right] = +\vert \boldsymbol{\psi} ,\vk \rangle	\;,
\end{equation}
since $\hat{U}^{\Theta} \cdot  (\hat{U}^{\Theta})^* = + \hat{1}$ for electronic orbital degrees of freedom, or 
\begin{equation}
\label{theta_square}
	\breve{\Theta}^{-\boldsymbol{k}} \cdot \left.\breve{\Theta}^{\boldsymbol{k}}\right.^* = + \breve{1}\;.
\end{equation}
%
If the system has TRS, then $\Theta$ and $\{g\vert \boldsymbol{\tau}_g\} \in \mathcal{G}$ commute, such that 
\begin{equation}
	\breve{\Theta}^{g\boldsymbol{k}}  \cdot \left(\breve{S}^{\boldsymbol{k}}(\{g\vert \boldsymbol{\tau}_g\}) \right)^* = \breve{S}^{-\boldsymbol{k}}(\{g\vert \boldsymbol{\tau}_g\}) \cdot 	\breve{\Theta}^{\boldsymbol{k}} \;,
\end{equation}
then at the time reversal invariant momenta (TRIMs), $\Theta\bar{\boldsymbol{k}}=-\bar{\boldsymbol{k}} = \bar{\boldsymbol{k}} - \boldsymbol{K}_{\Theta}$ with a little co-group containing $g$, we have 
\begin{equation}
	\Theta^{\bar{\boldsymbol{k}}}  \cdot \left(\Gamma^{\bar{\boldsymbol{k}}}_{\alpha}\left(\{g\vert \boldsymbol{\tau}_g\}\right)\right)^* \cdot \left(\Theta^{\bar{\boldsymbol{k}}}\right)^{-1} = \Gamma^{\bar{\boldsymbol{k}}}_{\alpha}\left(\{g\vert \boldsymbol{\tau}_g\}\right)  \;.
\end{equation}
Furthermore at TRIMs, we have $\Theta^{\boldsymbol{k}} = + \left(\Theta^{\boldsymbol{k}}\right)^T$, i.e. $\Theta^{\boldsymbol{k}}$ is symmetric. Therefore, if $\Gamma^{\bar{\boldsymbol{k}}}_{\alpha}\left(\{g\vert \boldsymbol{\tau}_g\}\right)$ is equivalent to its complex conjugate, it can be made real and no extra degeneracies are expected from TRS. If, on the contrary, $\Gamma^{\bar{\boldsymbol{k}}}_{\alpha}\left(\{g\vert \boldsymbol{\tau}_g\}\right)$ is complex, then TRS imposes an extra degeneracy of the two complex conjugate IRREPs, see for instance Ref.~\cite{Lax} for more details.

\subsection{Symmetries of the Hamiltonian}\label{Ham_sym}
We also write down explicitly the symmetries of the Hamiltonian in the symmetry-Bloch-l\"owdin basis. 

{Point group symmetry:}
%
	$\hat{U}_{\{g\vert \boldsymbol{\tau}_g\}} \cdot H(\boldsymbol{k}) \cdot \hat{U}^{\dagger}_{\{g\vert \boldsymbol{\tau}_g\}} = H(g\boldsymbol{k})$
%

{Reciprocal translation symmetry:} 
%
	$\hat{T}(\boldsymbol{K}) \cdot H(\boldsymbol{k}+\boldsymbol{K}) \cdot \hat{T}^{\dagger}(\boldsymbol{K}) = H(\boldsymbol{k}) $
%

{Time reversal symmetry:}
%
	$\hat{U}^{\Theta} \cdot H^*(\boldsymbol{k}) \cdot \hat{U}^{\Theta \dagger} = H(-\boldsymbol{k})$
%

\subsection{IRREPs at high-symmetry points and lines of the BZ}
We briefly summarize the IRREPs at the high-symmetry points and lines of the BZ following Refs.~\cite{BradCrack} or \cite{Bilbao}. It is well known that nonsymmorphic space groups exhibit extra degeneracies at the BZ boundary. Of peculiar importance for our discussion are the allowed (projective) IRREPs at the high-symmetry points and lines shown in Fig.~\ref{fig_SG19_BZ_SM}. 
There is a single 2D projective IRREP at the points $\{X,Y,Z\}$ (green), a single 2D projective IRREP at the lines $\overline{AB}$ with $A\in\{X,Y,Z\}$ and $B\in\{S,T,U\}$ (black) (without TRS they split into two distinct 1D projective IRREPs), a single 4D projective IRREP at the $R$-point (red) (without TRS, it splits into two identical 2D projective IRREPs), and there are two distinct 2D projective IRREPs at the points $\{S,T,U\}$ and at the lines $\{\overline{RS},\overline{RT},\overline{RU}\}$ (blue) (without TRS they split into four distinct 1D projective IRREPs). The remaining regions have only 1D IRREPs. 

In essence, it is the the single 4D projective IRREP at $R$ (with TRS) that readily leads to the minimum four-band connectivity of the band structure for SG19. However much more can be learned about the band structure topology that is strongly constrained by the group of spatial symmetries. 

\begin{figure}[t]
\centering
\begin{tabular}{cc} 
	\includegraphics[width=0.25\linewidth]{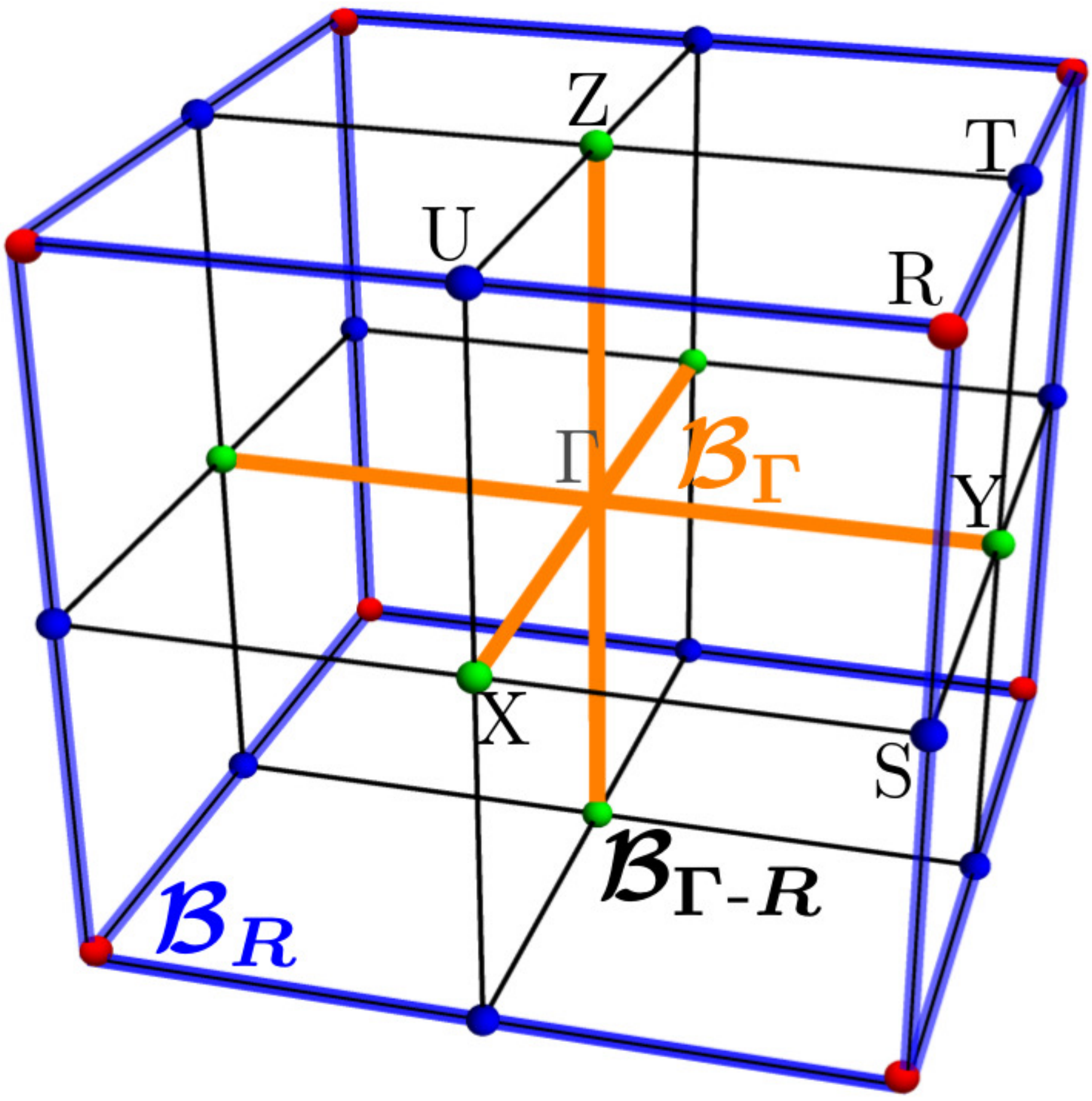} 
\end{tabular}
\caption{\label{fig_SG19_BZ_SM} The BZ with its high-symmetry regions for SG19.}
\end{figure}

\section{Smooth vs.~periodic gauge}\label{gauges}

We here introduce the smooth and periodic gauges, which are both used below and in the main text. 
The spectrum of the Hamiltonian operator is given through $\sigma(\mathcal{H}(\boldsymbol{k})) = \mathrm{eig}(H(\boldsymbol{k}))$, i.e.~the eigenvalues of the matrix $H(\boldsymbol{k})$. Anticipating the next section, the spectrum is invariant under reciprocal lattice translations, i.e. $\sigma(\mathcal{H}(\boldsymbol{k}+\boldsymbol{K})) = \sigma(\mathcal{H}(\boldsymbol{k}))$, and the eigenvectors are equal up to a global $U(4N)$ gauge transformation $\vert \boldsymbol{\psi}, \boldsymbol{k}+\boldsymbol{K}  \rangle \propto 	\vert \boldsymbol{\psi}, \boldsymbol{k}  \rangle$. 

We define the smooth gauge as the ordering and labeling of the eigenvectors, such that their corresponding eigenvalues are smooth as a function of $\boldsymbol{k}$ (i.e.~such that their derivative is continuous) over the different subspaces of the Brillouin zone ($\mathcal{B}_{\Gamma}$ and $\mathcal{B}_{R}$). More concretely, we label all the eigenvalues at $\Gamma$ (or the other high-symmetry points $U_i$) and then travel smoothly along each branch. In general, as we follow smoothly the $\alpha\equiv (\Gamma_j,n)$ branch, we have $\vert \psi^{(s)}_{\alpha} , \boldsymbol{k}+\boldsymbol{K} \rangle \neq  \vert \psi^{(s)}_{\alpha} , \boldsymbol{k} \rangle $ and $E^{(s)}_{\alpha}(\boldsymbol{k}+\boldsymbol{K}) \neq E^{(s)}_{\alpha}(\boldsymbol{k})$.

The periodic gauge is defined through
\begin{eqnarray}
	\vert \boldsymbol{\psi}^{(p)} , \boldsymbol{k}+\boldsymbol{K}  \rangle  &=& 	\mathrm{e}^{i \theta_{\boldsymbol{K}}} \vert \boldsymbol{\psi}^{(p)} , \boldsymbol{k}  \rangle\;,
\end{eqnarray}
i.e.~the eigenvectors are equal under a reciprocal lattice translation up to a global phase factor, which leads to 
\begin{eqnarray}
	E^{(p)}_{\alpha}(\boldsymbol{k}+\boldsymbol{K}) &=& E^{(p)}_{\alpha}(\boldsymbol{k}) \;.
\end{eqnarray}
In the following we drop the label $(p)$ (or $s$) and specify which gauge is assumed. 

We can always choose $\theta_{\boldsymbol{K}} = 0$ in the periodic gauge, in which case the following holds,
\begin{equation}
	\breve{U}(\boldsymbol{k}+\boldsymbol{K})	 = \hat{T}^{\dagger}(\boldsymbol{K}) \cdot \breve{U}(\boldsymbol{k})\;.
\end{equation}

Therefore from Eq.~(\ref{band_basis_K}) 
\begin{equation}
\label{K_shift_band_1}
\begin{aligned}
	\vert \boldsymbol{\psi} ,\boldsymbol{k}+\boldsymbol{K} \rangle &= \vert \boldsymbol{\psi} ,\boldsymbol{k} \rangle \cdot \breve{M}_{\boldsymbol{k}}(\boldsymbol{K}) = \vert \boldsymbol{\psi} ,\boldsymbol{k} \rangle  \;,
\end{aligned}
\end{equation}
since
\begin{equation}
\label{K_shift_band_2}
\begin{aligned}
	\breve{M}_{\boldsymbol{k}}(\boldsymbol{K}) &= \breve{U}^{\dagger}(\boldsymbol{k}) \cdot \hat{T}(\boldsymbol{K}) \cdot \breve{U}(\boldsymbol{k}+\boldsymbol{K}) \\
	&= \breve{U}^{\dagger}(\boldsymbol{k}) \cdot \hat{T}(\boldsymbol{K}) \cdot  \hat{T}^{\dagger}(\boldsymbol{K}) \cdot \breve{U}(\boldsymbol{k}) \\
	&=  \boldsymbol{1}   \;.
\end{aligned}
\end{equation}

At a high-symmetry point located on the boundary of the BZ, i.e. $g\boldsymbol{k}^* = \boldsymbol{k}^* - \boldsymbol{K}_g$, the periodic gauge leads to the following convenient simplification of the band representation of the symmetry operator $\{g\vert \boldsymbol{\tau}_g\}$, 
\begin{eqnarray}
	\breve{S}^{\boldsymbol{k}^*}\left(\{g\vert \boldsymbol{\tau}_g\}\right) &=& \mathrm{e}^{-i g\boldsymbol{k}^*\cdot \boldsymbol{\tau}_g}  \cdot \left[ \breve{U}^{\dagger}(\boldsymbol{k}^*) \cdot \hat{T}(-\boldsymbol{K}_g) \cdot \hat{U}_{\{g\vert \boldsymbol{\tau}_g\}} \cdot \breve{U}(\boldsymbol{k}^*) \right]\;. 
\end{eqnarray}

\section{Band permutations and connectivity}

In this section we show in detail how the nonsymmorphic space group lead to unavoidable band permutations along special directions, both within the BZ and at its boundaries, which results in a nontrivial connectivity of of any four band subspace.

\subsection{Proof of the connectivity for an isolated four-band subspace}

We start with a $4$-band subspace isolated in energy, i.e.~separated from other bands by an energy gap both above and below.

\subsubsection{High-symmetry lines within the BZ, $\mathcal{B}_{\Gamma}$}

Michel and Zak early derived the symmetry enforced band permutations over $\mathcal{B}_{\Gamma}$ for SG19 from an argument strictly based on IRREPs \cite{Zak2}. Here we derive the same result in a more detailed way by referring to a generic basis set and Hamiltonian both constructed explicitly from symmetry. This allows us to reveal the machinery hidden under the representation arguments.

We start by showing that, as a consequence of the three screw axis symmetries and reciprocal translational symmetry, any $4$-band subspace isolated in energy must satisfy: (i) the IRREPs at the $\Gamma$-point are $\{\Gamma^{\boldsymbol{0}}_1,\Gamma^{\boldsymbol{0}}_2,\Gamma^{\boldsymbol{0}}_3,\Gamma^{\boldsymbol{0}}_4\}$ and (ii) the four bands must be connected over the section of the BZ $\mathcal{B}_{\Gamma} = \cup_{X_i = X,Y,Z} \overline{\Gamma X_i}$ (the three high-symmetry lines crossing $\Gamma$), i.e.~there can be no energy gap within this section of the BZ separating a subset of the four bands from the remaining bands. A direct consequence of this four-band connectivity is the existence of two (simple) Dirac points within the BZ located on one of the lines of $\mathcal{B}_{\Gamma}$. 

Since only 1D IRREPs are allowed at the $\Gamma$-point we can write the symmetry-Bloch basis of the four-band subspace as $\vert \boldsymbol{\phi}^{sub} ,\vk \rangle = \left( \vert \phi^{sub}_{\Gamma_{j_1}} ,\vk \rangle,\vert \phi^{sub}_{\Gamma_{j_2}} ,\vk \rangle,\vert \phi^{sub}_{\Gamma_{j_3}} ,\vk \rangle,\vert \phi^{sub}_{\Gamma_{j_4}} ,\vk \rangle\right)^T $ and choose a gauge such that it matches with the band basis at the $\Gamma$-point, i.e. 
\begin{equation}
	\vert \boldsymbol{\psi}^{sub} ,\boldsymbol{0} \rangle = \vert \boldsymbol{\phi}^{sub} ,\boldsymbol{0} \rangle \;. 
\end{equation}
Note that the basis of an isolated four-band subspace can in principle be obtained through a partial diagonalization of the Hamiltonian written in the symmetry-Bloch-L\"owdin basis, i.e.~there is an unitary transformation
\begin{eqnarray}
		\left( \begin{array}{c} \vert \boldsymbol{\phi}^{sub} ,\vk \rangle \\
		\vert \boldsymbol{\phi}^{\backslash sub} ,\vk\rangle  \end{array}\right)^T 
		&=& \vert \boldsymbol{\phi} ,\vk \rangle \cdot \tilde{U}(\vk )	\;,\\
		 \tilde{U}(\vk )&=& \left( \begin{array}{cc} \tilde{U}^{sub}(\vk ) & \tilde{U}^{\backslash sub}(\vk )  \end{array}\right) \;, \nonumber
\end{eqnarray}
where $\backslash sub \equiv \mathcal{H}\backslash sub$ means the complement of the four-band subspace within the whole Hilbert space $\mathcal{H}$. The band basis is then given through the unitary transformation that diagonalizes $H^{sub}(\vk)$, i.e. through $\vert \boldsymbol{\psi}^{sub} ,\boldsymbol{k} \rangle = \vert \boldsymbol{\phi}^{sub} ,\boldsymbol{k} \rangle \cdot \breve{U}^{sub}(\boldsymbol{k})$. In the following we choose a gauge such that the labeled eigenvalues of the four-band subspace, $\{E_{\Gamma_{j_1}}(\vk),E_{\Gamma_{j_2}}(\vk),E_{\Gamma_{j_3}}(\vk),E_{\Gamma_{j_4}}(\vk)\}$, are smooth over the whole BZ (the labeling refers to the IRREPs at the $\Gamma$-point), i.e. we assume the smooth gauge. Note that in general in the smooth gauge $E_{j_1}(\vk) \neq E_{j_1}(\vk+\boldsymbol{b}_{1,2,3}) $.

The symmetry transformations of $\vert \boldsymbol{\phi}^{sub} ,\vk \rangle$ and $\vert \boldsymbol{\psi}^{sub} ,\vk \rangle$ are known from Eqs.~(\ref{sym_basis_trans}) and (\ref{band_basis_trans}). The Hamiltonian written in the symmetry-Bloch basis is block diagonal along the high-symmetry lines $\vk^* \in \overline{\Gamma X_i}$, which means that the blocks connect only components $\vert \phi^{sub}_{\Gamma_j} ,\vk^* \rangle $ that have the same compatibility relation from the $\Gamma$-point to the line, i.e.~$\forall \Gamma_{j} ~\vert~\Gamma^{\boldsymbol{0}}_{j} \rightarrow \Gamma^{\boldsymbol{k}^*}_{j'}$. Therefore, the unitary matrix $\breve{U}^{sub}( \vk^*)$ also has a block-diagonal form ($\breve{U}^{sub}_{j_a\rightarrow j', j_b\rightarrow j''}( \vk^*) \propto \delta_{j',j''}$):
\begin{equation}
\begin{aligned}
	\vert \psi^{sub}_{\Gamma_{j\rightarrow j'}} ,\boldsymbol{k}^* \rangle &= \sum\limits_{j_a} \vert \phi^{sub}_{\Gamma_{j_a \rightarrow j''}} ,\boldsymbol{k}^* \rangle  \breve{U}^{sub}_{j_a\rightarrow j'', j\rightarrow j'}(\boldsymbol{k}^*)  \\
	&= \sum\limits_{j_a} \vert \phi^{sub}_{\Gamma_{j_a \rightarrow j'}} ,\boldsymbol{k}^* \rangle  \breve{U}^{sub}_{j_a\rightarrow j', j\rightarrow j'}(\boldsymbol{k}^*)  \;,
\end{aligned}
\end{equation}
where we sum only over the components $j_a$ that maps to $j'$ determined by the compatibility relation $\Gamma^{\boldsymbol{0}}_{j} \rightarrow \Gamma^{\boldsymbol{k}^*}_{j'}$. Note that there are only 1D IRREPs on the high-symmetry lines within the BZ. Then using Eqs.~(\ref{sym_basis_trans}) and (\ref{band_basis_trans}), and the fact that $C_{2i} \vk^* =\vk^*$ if $\vk^* \in \overline{\Gamma X_i}$, we find
\begin{equation}
\label{sym_band_u_g}
\begin{aligned}
	^{\{C_{2i}\vert \tau_i\}} \vert \psi^{sub}_{\Gamma_{j\rightarrow j'}} ,\vk^* \rangle &= 
	\mathrm{e}^{- i \boldsymbol{k}^*\cdot \boldsymbol{\tau}_g} \sum\limits_{j_a,j_b,j_c} \vert \psi^{sub}_{\Gamma_{j_a\rightarrow j''}} , \vk^* \rangle  [\breve{U}^{sub\dagger}(\vk^*)]_{j_a,j_b} [\hat{U}^{sub}_{\{g\vert \boldsymbol{\tau}_g\}}]_{j_b,j_c}  [\breve{U}^{sub}( \vk^*)]_{j_c,j} \\
	&= 
	\chi^{\Gamma_{j'}}(C_{2i}) \mathrm{e}^{- i \boldsymbol{k}^*\cdot \boldsymbol{\tau}_g} \sum\limits_{j_a} \vert \psi^{sub}_{\Gamma_{j_a\rightarrow j''}} , \vk^* \rangle \sum\limits_{j_b}  [\breve{U}^{sub\dagger}(\vk^*)]_{j_a,j_b}   [\breve{U}^{sub}( \vk^*)]_{j_b,j} \\
	&= 
	\chi^{\Gamma_{j'}}(C_{2i}) \mathrm{e}^{- i \boldsymbol{k}^*\cdot \boldsymbol{\tau}_g} \vert \psi^{sub}_{\Gamma_{j\rightarrow j'}} , \vk^* \rangle  \;,
\end{aligned}
\end{equation}
where we have used $[\hat{U}^{sub}_{\{C_{2i}\vert \boldsymbol{\tau}_i\}}]_{j_b\rightarrow j'',j_c\rightarrow j'} = \delta_{j_b,j_c} \delta_{j'',j'} \chi^{\Gamma_{j'}}(C_{2i})$ in the second line and the orthogonality relation $\sum_{j_b}[\breve{U}^{sub\dagger}(\vk^*)]_{j_a,j_b}    [\breve{U}^{sub}( \vk^*)]_{j_b,j} = \delta_{j_a,j}$ in the third line. 

Furthermore, the symmetry-Bloch-L\"owdin basis of the whole Hilbert space $\mathcal{H}$ transforms under a reciprocal translation as 
\begin{eqnarray}
\label{tran_sub_1}
		\left[\begin{array}{c} \vert \boldsymbol{\phi}^{sub} ,\vk+\boldsymbol{K} \rangle \\
		\vert \boldsymbol{\phi}^{\backslash sub} ,\vk+\boldsymbol{K} \rangle  \end{array}\right]^T 
		&=& \left[ \begin{array}{c} \vert \boldsymbol{\phi}^{sub} ,\vk \rangle \\
		\vert \boldsymbol{\phi}^{\backslash sub} ,\vk \rangle  \end{array}\right]^T	\cdot \tilde{T}_{\boldsymbol{k}}(\boldsymbol{K})  \;, \\
		\tilde{T}_{\boldsymbol{k}}(\boldsymbol{K})  &=& \tilde{U}^{\dagger}(\vk) \cdot \hat{T}(\boldsymbol{K}) \cdot \tilde{U}(\vk + \boldsymbol{K}) \;, \nonumber
\end{eqnarray}
where the unitary translation matrix in the symmetry-Bloch-L\"owdin basis, i.e. $ \hat{T}(\boldsymbol{K})$, is known explicitly (through the symmetry decomposition of the degrees of freedom of the system, see Section \ref{4N_dof}). First of all, $\tilde{T}_{\boldsymbol{k}}(\boldsymbol{K}) $ is unitary and since we assumed that the $4N$-band spectrum was fully gapped between the four-band subspace $\mathcal{H}^{sub}$ and its complement $\mathcal{H}^{\backslash sub}$ it must have one block diagonal form, each block acting separately on the two subspaces:
\begin{equation}
\label{tran_sub_2}
	\tilde{T}_{\boldsymbol{k}}(\boldsymbol{K}) = \left(\begin{array}{cc} \tilde{T}^{sub}_{\boldsymbol{k}}(\boldsymbol{K}) & 0 \\
	0 & \tilde{T}^{\backslash sub}_{\boldsymbol{k}}(\boldsymbol{K}) \end{array}\right)	\;.
\end{equation}
From Eqs.~(\ref{tran_sub_1}) and (\ref{tran_sub_2}) we have $\vert \boldsymbol{\phi}^{sub} ,\vk+\boldsymbol{K} \rangle = \vert \boldsymbol{\phi}^{sub} ,\vk \rangle \cdot  \tilde{T}^{sub}_{\boldsymbol{k}}(\boldsymbol{K})$ with $ \tilde{T}^{sub}_{\boldsymbol{k}}(\boldsymbol{K})$ unitary. Furthermore the band basis follows 
\begin{equation}
\begin{aligned}
	\vert \boldsymbol{\psi}^{sub} ,\boldsymbol{0}+\boldsymbol{K} \rangle &= \vert \boldsymbol{\psi}^{sub} ,\boldsymbol{0} \rangle \cdot \breve{M}^{sub}_{\boldsymbol{0}}(\boldsymbol{K}) \;,\\
 	\breve{M}^{sub}_{\boldsymbol{0}}(\boldsymbol{K}) &= \tilde{T}^{sub}_{\boldsymbol{0}}(\boldsymbol{K}) \cdot \breve{U}^{sub}(\boldsymbol{K})
\end{aligned}
\end{equation}
where $  \breve{M}^{sub}_{\boldsymbol{0}}(\boldsymbol{K}) $ is unitary. We then conclude that the set of energy eigenvalues at $\vk$ and $\vk+\boldsymbol{K}$ must be identical, i.e.~$\sigma\left( \mathcal{H}^{sub}(\boldsymbol{k}+\boldsymbol{K})\right) = \sigma\left( \mathcal{H}^{sub}(\boldsymbol{k})\right)$. Therefore, within the smooth gauge, $\breve{M}^{sub}_{\boldsymbol{0}}(\boldsymbol{K})$ is given by a permutation of the components of the band basis (up to a global phase factor that can be gauged away) and, correspondingly, there is a permutation of the (labeled) eigenvalue-branches, i.e.~bands. We write this as $\mathcal{P}_{\overline{\Gamma X_i}}$.

We finally derive the constraints on the set of IRREPs $\{\Gamma_{j_1},\Gamma_{j_2},\Gamma_{j_3},\Gamma_{j_4}\}$ and the permutations $\mathcal{P}_{\overline{\Gamma X_i}}$ due to the group of symmetries. Let us start assuming that the IRREP $\Gamma_1$ is present in the four-band subspace at the $\Gamma$-point. Then by Eq.~(\ref{sym_band_u_g}) we have that $\vert \psi^{sub}_{\Gamma_1} , \boldsymbol{b}_1\rangle $ has a symmetry eigenvalue 
\begin{equation}
	\lambda^{\Gamma_1}_{C_{2x}} (\boldsymbol{b}_1) = -\lambda^{\Gamma_1}_{C_{2x}} (\boldsymbol{0})\;.
\end{equation}
This means that the component $\vert u^{sub}_{\Gamma_1} , \boldsymbol{0}\rangle $ is mapped to $\vert u^{sub}_{\Gamma_2} , \boldsymbol{0}\rangle $ or $\vert u^{sub}_{\Gamma_3} , \boldsymbol{0}\rangle $ after a shift by $\boldsymbol{b}_1$. Exhausting all the symmetries of $D_2$ and the shifts in the three directions $\{\boldsymbol{b}_i\}_{i=1,2,3}$, it is straightforward to conclude that such an isolated four-band subspace must (i) be composed of the components $\{\Gamma_1,\Gamma_2,\Gamma_3,\Gamma_4\}$ at the $\Gamma$-point and (ii) realize one set of three band permutations among those listed in Table I in the main text. 

We note that each band permutation of Table I form a representation of the Klein four-group $\mathbb{Z}_2 \times \mathbb{Z}_2$. Also, any four-band model of SG19 must realize at least two distinct permutations from Table I leading to the stated connectivity of the four bands over $\mathcal{B}_{\Gamma}$. It is worth noting here that we did not rely on TRS in this derivation.

\subsubsection{High-symmetry lines on the BZ boundary, $\mathcal{B}_{R}$}\label{projective_reps}

Having studied the high-symmetry lines inside the BZ above we now turn to the high-symmetry lines on the BZ boundary.
It is well established that the nonsymmorphicity leads to the need of projective representations of the little co-groups \cite{BradCrack}. We already mentioned the single 2D projective IRREP at the points $\{X,Y,Z\}$, the single 4D projective IRREP at the $R$-point (assuming TRS), and the single 2D projective IRREP on the lines $\overline{AB}$ for $A\in\{X,Y,Z\}$ and $B\in\{S,T,U\}$ (assuming TRS) [Fig.~\ref{fig_SG19_BZ_SM}]. Since each of those high-symmetry regions only allows a single IRREP, no extra symmetry protected band-crossing can occur on them. Therefore we only need to consider the lines $\{\overline{RS},\overline{RT},\overline{RU}\}$ and the points $S,T,U$, which allow the realization of two distinct 2D projective IRREPs (assuming TRS) \cite{BradCrack}. 

At a high-symmetry point $\boldsymbol{k}^*$ where $\boldsymbol{k}^* = g \boldsymbol{k}^* + \boldsymbol{K}_g $ and $g \in \overline{\mathcal{G}}^{ \boldsymbol{k}^* }$ (the little co-group), the projective IRREPs take the general form
\begin{equation}
	\Gamma^{\boldsymbol{k}^*}_{\alpha} \left(\{g\vert \boldsymbol{\tau}_g\}\right) = \mathrm{e}^{-i \boldsymbol{k}^* \cdot \boldsymbol{\tau}_g } D^{\boldsymbol{k}^*}_{\alpha} \left(\{g\vert \boldsymbol{\tau}_g\}\right)	\;,
\end{equation}
with the matrix algebra 
\begin{equation}
	D^{\boldsymbol{k}^*}_{\alpha} \left(\{g_1\vert \boldsymbol{\tau}_{g_1}\}\right) 
	D^{\boldsymbol{k}^*}_{\alpha} \left(\{g_2\vert \boldsymbol{\tau}_{g_2}\}\right) = 
	\mathrm{e}^{-i \boldsymbol{K}_{g_1} \cdot  \boldsymbol{\tau}_{g_2}} D^{\boldsymbol{k}^*}_{\alpha} \left(\{g_3\vert \boldsymbol{\tau}_{g_3}\}\right)	\;,
\end{equation}
The factor $\mu(g_1,g_2) = \mathrm{e}^{-i \boldsymbol{K}_{g_1} \cdot  \boldsymbol{\tau}_{g_2}}$ defines the factor system of the projective representation $\alpha$ \cite{BradCrack}. Shifting by a reciprocal lattice vector $\boldsymbol{K}$, we get
\begin{equation}
	\Gamma^{\boldsymbol{k}^*+\boldsymbol{K}}_{\alpha} \left(\{g\vert \boldsymbol{\tau}_g\}\right) = \mathrm{e}^{-i \boldsymbol{K} \cdot \boldsymbol{\tau}_g }  \left[ \mathrm{e}^{-i \boldsymbol{k}^* \cdot \boldsymbol{\tau}_g } D^{\boldsymbol{k}^*+\boldsymbol{K} }_{\alpha} \left(\{g\vert \boldsymbol{\tau}_g\}\right) \right]	\;,
\end{equation}
with the matrix algebra 
\begin{equation}
	D^{\boldsymbol{k}^*+\boldsymbol{K}}_{\alpha} \left(\{g_1\vert \boldsymbol{\tau}_{g_1}\}\right) 
	D^{\boldsymbol{k}^*+\boldsymbol{K}}_{\alpha} \left(\{g_2\vert \boldsymbol{\tau}_{g_2}\}\right) = 
	\mathrm{e}^{ -i [\boldsymbol{K} - g_1 \boldsymbol{K}] } \mathrm{e}^{-i \boldsymbol{K}_{g_1} \cdot  \boldsymbol{\tau}_{g_2}} D^{\boldsymbol{k}^*+\boldsymbol{K}}_{\alpha} \left(\{g_3\vert \boldsymbol{\tau}_{g_3}\}\right)	\;,
\end{equation}
with a new factor system of the shifted projective representation
\begin{equation}
	\nu(g_1,g_2) = \mathrm{e}^{ -i [\boldsymbol{K} - g_1 \boldsymbol{K}] \cdot \boldsymbol{\tau}_{g_2} } \mu(g_1,g_2) \;.
\end{equation} 
For such a transformation the two factor systems, $\mu$ and $\nu$, belong to the same equivalence class, but it might be that their corresponding matrix representations, i.e.~$D^{\boldsymbol{k}^*}_{\alpha}$ and $D^{\boldsymbol{k}^*+\boldsymbol{K}}_{\alpha}$, are rotated and not identically the same. However, for the particular case we are considering, take for instance the line $\boldsymbol{k}^* \in \overline{RS}$ and $\boldsymbol{K} = \boldsymbol{b}_3$, we trivially find $\nu(g_1,g_2) = \mu(g_1,g_2)$ for all $g_1,g_2 \in \overline{\boldsymbol{\mathcal{G}}}^{ \boldsymbol{k}^* } = \{ E,C_{2z} \} $, and similarly for the lines $\overline{RT}$ and $\overline{RU}$. We then conclude the following relationship between shifted projective IRREPs, assuming $g \boldsymbol{b}_i = \boldsymbol{b}_i$,
\begin{equation}
\label{eq_proj_IRREP}
	\Gamma^{\boldsymbol{k}^*+\boldsymbol{b}_i}_{\alpha} \left(\{g\vert \boldsymbol{\tau}_g\}\right) = \mathrm{e}^{-i \boldsymbol{b}_i \cdot \boldsymbol{\tau}_g } \Gamma^{\boldsymbol{k}^*}_{\alpha} \left(\{g\vert \boldsymbol{\tau}_g\}\right)	\;.
\end{equation}
Along the line $\overline{RS}$, we then take $g=C_{2z}$ and $\boldsymbol{b}_3$, which gives $\mathrm{e}^{-i \boldsymbol{b}_3 \cdot \boldsymbol{\tau}_{z} } = -1$. We get an identical result for $\overline{RT}$ where we take $(g,\boldsymbol{K} )=(C_{2x}, \boldsymbol{b}_1) $, and $\overline{RU}$ where we take $(g,\boldsymbol{K} )=(C_{2y}, \boldsymbol{b}_2) $. 

The two 2D projective IRREPs at $\boldsymbol{k}^* \in \overline{RS}$ (assuming TRS) \cite{BradCrack}, written here as $\{\Gamma_5,\Gamma_6\}$, can be defined through their characters
\begin{equation}
\left\{
\begin{aligned}
	\chi^{\boldsymbol{k}^*}_{5}\left(\{C_{2z}\vert \boldsymbol{\tau}_{z}\}\right) = \mathrm{tr}~ \Gamma^{\boldsymbol{k}^*}_{5} \left(\{C_{2z}\vert \boldsymbol{\tau}_{z}\}\right)  &= + 2i\mathrm{e}^{-i \boldsymbol{k}^*\cdot \boldsymbol{\tau}_z} \\
	\chi^{\boldsymbol{k}^*}_{6}\left(\{C_{2z}\vert \boldsymbol{\tau}_{z}\}\right) =\mathrm{tr}~ \Gamma^{\boldsymbol{k}^*}_{6} \left(\{C_{2z}\vert \boldsymbol{\tau}_{z}\}\right)  &= - 2i\mathrm{e}^{-i \boldsymbol{k}^*\cdot \boldsymbol{\tau}_z}
\end{aligned}
\right. \;,
\end{equation}
such that 
\begin{equation}
\left\{
\begin{aligned}
	\chi^{S}_{5}\left(\{C_{2z}\vert \boldsymbol{\tau}_{z}\}\right)  &= + 2 \\
	\chi^{S}_{6}\left(\{C_{2z}\vert \boldsymbol{\tau}_{z}\}\right)  &= - 2
\end{aligned}
\right. \;.
\end{equation}
Therefore, with Eq.~(\ref{eq_proj_IRREP}) we find
\begin{equation}
\begin{aligned}
	\Gamma^{S+\boldsymbol{b}_3}_{5} \left(\{C_{2z}\vert \boldsymbol{\tau}_{z}\}\right) &= (-1) \cdot \Gamma^{S}_{5} \left(\{C_{2z}\vert \boldsymbol{\tau}_{z}\}\right) \\
	&= \Gamma^{S}_{6} \left(\{C_{2z}\vert \boldsymbol{\tau}_{z}\}\right) \;.
\end{aligned}
\end{equation}
Assuming the smooth gauge and following a similar argument as in the previous section, the two twofold degenerate eigenvalues $E_5(S)$ and $E_6(S)$ must be permuted at $S+\boldsymbol{b}_3$, when we smoothly follow the branches $\{E_{5}(\boldsymbol{k}^*),E_{6}(\boldsymbol{k}^*) \}$ along $\overline{RS}$, i.e. 
\begin{equation}
\label{permutation_BZB}
	\left(\begin{array}{c} E_5(S+\boldsymbol{b}_3)\\
	E_6(S+\boldsymbol{b}_3) \end{array}\right) = \mathcal{P}_{\overline{RS}} \left(\begin{array}{c} E_5(S)\\
	E_6(S) \end{array}\right)  = \left(\begin{array}{c} E_6(S)\\
	E_5(S) \end{array}\right)\;,
\end{equation}
where we write $\mathcal{P}_{\overline{RS}} = (56)$.
Hence the two branches must cross over the line $\overline{RS}$ (and similarly for $\overline{RT}$ and $\overline{RU}$) leading to a symmetry protected double Dirac point. Taking the point group $D_2$ into account, we naturally find that the crossing point lies at $R$ (strictly speaking, the permutation over a high-symmetry line imposes an odd number of band crossings which can be adiabatically mapped into a single band crossing, which then has to lie at the $D_2$ symmetric $R$-point). 

\subsection{Proof of the connectivity for an explicit four-band model}

After providing the general proof of the four-band connectivity, let us also show it for an explicit four-band model in SG19. Here we only consider the $\mathcal{B}_{\Gamma}$ subspace, the argument works similarly (but with projective IRREPs as in the previous section) for $\mathcal{B}_{R}$. 

For this we set the four representative sub-lattice sites to
\begin{equation}
\label{explicit_4B}
\left\{
\begin{aligned}
	\boldsymbol{r}_1 & = \alpha \boldsymbol{a}_1 + \beta \boldsymbol{a}_2 + \gamma \boldsymbol{a}_3\\
	\boldsymbol{r}_2 & = \left(\dfrac{1}{2}+\alpha\right) \boldsymbol{a}_1 + \left(\dfrac{1}{2}-\beta\right) \boldsymbol{a}_2 - \gamma \boldsymbol{a}_3\\
	\boldsymbol{r}_3 & = -\alpha \boldsymbol{a}_1 +  \left(\dfrac{1}{2}+\beta\right) \boldsymbol{a}_2 +  \left(\dfrac{1}{2}-\gamma\right) \boldsymbol{a}_3\\
	\boldsymbol{r}_4 & =  \left(\dfrac{1}{2}-\alpha\right) \boldsymbol{a}_1 - \beta \boldsymbol{a}_2 + \left(\dfrac{1}{2}+\gamma\right) \boldsymbol{a}_3
\end{aligned}\right.
\end{equation}
Let us first focus on the line $k_x\in \overline{\Gamma X}$. Because of the screw symmetry $\{C_{2x}\vert \boldsymbol{\tau}_x\}$ the four-band Hamiltonian in the symmetry-Bloch-L\"owdin basis splits into two blocks as
\begin{eqnarray}
\label{H_4B_blocks}
	H(k_x) &=& \left(\begin{array}{cc} H^{(\Gamma_1,\Gamma_4)}(k_x) & 0 \\
	0 & H^{(\Gamma_2,\Gamma_3)}(k_x) \end{array}\right) \;,
\end{eqnarray}
with 
\begin{equation}
\label{H_4B_blocks_b}
\begin{aligned}
	H^{(\Gamma_1,\Gamma_4)}(k_x) &= \left(\begin{array}{cc} h^{(\Gamma_1)}_{\Gamma_1\Gamma_1}(k_x) & h^{(\Gamma_4)}_{\Gamma_1\Gamma_4}(k_x) \\
	h^{(\Gamma_4)*}_{\Gamma_1\Gamma_4}(k_x) & h^{(\Gamma_1)}_{\Gamma_4\Gamma_4}(k_x) \end{array}\right) \;,\\
	H^{(\Gamma_2,\Gamma_3)}(k_x) &= \left(\begin{array}{cc} h^{(\Gamma_1)}_{\Gamma_2\Gamma_2}(k_x) & h^{(\Gamma_4)}_{\Gamma_2\Gamma_3}(k_x) \\
	h^{(\Gamma_4)*}_{\Gamma_2\Gamma_3}(k_x) & h^{(\Gamma_1)}_{\Gamma_3\Gamma_3}(k_x) \end{array}\right) \;,
\end{aligned}
\end{equation}
since only the terms that are basis functions of $\Gamma_1$ and $\Gamma_4$ are even under $C_{2x}$.
In the symmetry basis, the operator for a reciprocal translation by $\boldsymbol{b}_1$ is explicitly given by
\begin{equation}
\begin{aligned}
	\hat{T}(\boldsymbol{b}_1) &= \left(\begin{array}{cccc} 0 & T_1 \\
		T_1 & 0
	\end{array}\right)\;,\\
	T_1 &= \left(\begin{array}{cc}  \cos 2\pi \alpha  & i  \sin 2\pi \alpha \\
		 i\sin 2\pi \alpha  &  \cos 2\pi \alpha 
	\end{array}\right)\;.
\end{aligned}
\end{equation}
and for a reciprocal translation by $2\boldsymbol{b}_1$ it is given by
\begin{equation}
\begin{aligned}
	\hat{T}(2\boldsymbol{b}_1) &= \left(\begin{array}{cccc}  T_2 & 0  \\
		  0 &T_2  
	\end{array}\right)\;,\\
	T_2 &=  \left(\begin{array}{cccc}  \cos 4\pi \alpha  & i  \sin 4\pi \alpha \\
		 i\sin 4\pi \alpha  &  \cos 4\pi \alpha  
	\end{array}\right) \;.
\end{aligned}
\end{equation}
By the symmetry of the Hamiltonian under reciprocal translation we have
\begin{equation}
	 H(k_x+2\boldsymbol{b}_1) = \hat{T}^{\dagger}(2\boldsymbol{b}_1)\cdot H(k_x) \cdot \hat{T}(2\boldsymbol{b}_1)\;,
\end{equation}
which for the blocks in Eq.~(\ref{H_4B_blocks}) means
\begin{equation}
\label{T2_constraint}
\begin{aligned}
	H^{(\Gamma_1,\Gamma_4)}(k_x+2\boldsymbol{b}_1) &= T^{\dagger}_2 \cdot H^{(\Gamma_1,\Gamma_4)}(k_x) \cdot T_2 \;,\\
	H^{(\Gamma_2,\Gamma_3)}(k_x+2\boldsymbol{b}_1) &= T^{\dagger}_2 \cdot H^{(\Gamma_2,\Gamma_3)}(k_x) \cdot T_2 \;.
\end{aligned}
\end{equation}
Going to the band basis, we have 
\begin{equation}
	\left.\breve{U}^{(\Gamma_1,\Gamma_4)}\right.^{\dagger}(k_x) \cdot  H^{(\Gamma_1,\Gamma_4)}(k_x) \cdot \breve{U}^{(\Gamma_1,\Gamma_4)}(k_x) = 
	\left(\begin{array}{cc} E_1(k_x) & 0 \\
		0 & E_4(k_x)
	 \end{array}\right)\;,
\end{equation}
and similarly for the block $(\Gamma_2,\Gamma_3)$. But from Eq.~(\ref{T2_constraint}) we also have
\begin{equation}
	\left(\begin{array}{cc} E_1(k_x) & 0 \\
		0 & E_4(k_x)
	 \end{array}\right) = \left[T^{\dagger}_2 \breve{U}^{(\Gamma_1,\Gamma_4)}(k_x)  \right]^{\dagger}\cdot  H^{(\Gamma_1,\Gamma_4)}(k_x+2\boldsymbol{b}_1) \cdot \left[T^{\dagger}_2 \breve{U}^{(\Gamma_1,\Gamma_4)}(k_x)  \right]\;,
\end{equation}
such that either $E_1(k_x+2\boldsymbol{b}_1) = E_1(k_x)$ or $E_1(k_x+2\boldsymbol{b}_1) = E_4(k_x)$. The second case is excluded since the two branches $E_1(k_x)$ and $E_4(k_x)$ are both even under $C_{2x}$ (they have the same compatibility relations along the axis $\overline{\Gamma X}$) and they can hybridize, therefore these two branches cannot permute (and cross each other) along $\Gamma -(2\boldsymbol{b}_1)$.
Similarly we have
\begin{equation}
	 H(k_x+\boldsymbol{b}_1) = \hat{T}^{\dagger}(\boldsymbol{b}_1)\cdot H(k_x) \cdot \hat{T}(\boldsymbol{b}_1)\;,
\end{equation}
which for the blocks in Eq.~(\ref{H_4B_blocks}) means
\begin{eqnarray}
	H^{(\Gamma_1,\Gamma_4)}(k_x+\boldsymbol{b}_1) &=& T^{\dagger}_1 \cdot H^{(\Gamma_2,\Gamma_3)}(k_x) \cdot T_1 \;,\\
	H^{(\Gamma_2,\Gamma_3)}(k_x+\boldsymbol{b}_1) &=& T^{\dagger}_1 \cdot H^{(\Gamma_1,\Gamma_4)}(k_x) \cdot T_1 \;.
\end{eqnarray}
Therefore, a translation by $\boldsymbol{b}_1$ connects the two blocks in Eq.~(\ref{H_4B_blocks}) through a unitary transformation. This leads to the following constraint on the band spectrum, $\sigma \left( H^{(\Gamma_1,\Gamma_4)}(k_x+\boldsymbol{b}_1) \right) = \sigma \left( H^{(\Gamma_2,\Gamma_3)}(k_x) \right)$. 

Going to the band basis, we derive 
\begin{equation}
	\left(\begin{array}{cc} E_2(k_x) & 0 \\
		0 & E_3(k_x)
	 \end{array}\right) = \left[T^{\dagger}_1 \breve{U}^{(\Gamma_2,\Gamma_3)}(k_x)  \right]^{\dagger}\cdot  H^{(\Gamma_1,\Gamma_4)}(k_x+\boldsymbol{b}_1) \cdot \left[T^{\dagger}_1 \breve{U}^{(\Gamma_2,\Gamma_3)}(k_x)  \right]\;,
\end{equation}
from which we conclude that either
\begin{equation}
\left\{
\begin{aligned}
	E_1(k_x+\boldsymbol{b}_1) &= E_2(k_x) \\
	E_4(k_x+\boldsymbol{b}_1) &= E_3(k_x)
\end{aligned}\right. \;,
\end{equation}
in which case we have the permutation rule $\mathcal{P}_{\overline{\Gamma \boldsymbol{b}_1}} = (12)(34)$, or 
\begin{equation}
\left\{
\begin{aligned}
	E_1(k_x+\boldsymbol{b}_1) &= E_3(k_x) \\
	E_4(k_x+\boldsymbol{b}_1) &= E_2(k_x)
\end{aligned}\right. \;,
\end{equation}
in which case we have the permutation rule $\mathcal{P}_{\overline{\Gamma \boldsymbol{b}_1}} = (13)(24)$. Proceeding similarly in the two other directions, this proves in a more explicit manner the permutation rules of Table I.

\subsection{Band structures for four- and eight-band tight-binding models}

We here describe briefly how we constructed the four-band and eight-band tight-binding Hamiltonian used to generate the band structures of the main text (see Figs.~1(c) and 3(a-b)). In both cases we assume only one $s$-electronic orbital per site. We start from the four representative sub-lattice sites of Eq.~(\ref{explicit_4B}) and write the Hamiltonian in the site basis Eq.~(\ref{site_basis}) with the Hamiltonian matrix elements $\vert \varphi_{\boldsymbol{r}_i ,\boldsymbol{k}} \rangle \langle \varphi_{\boldsymbol{r}_j  , \boldsymbol{k}} \vert$ ($i,j=1,...,4$) given by $h_{ij}(\boldsymbol{k}) = t_{ij} \mathrm{e}^{i \boldsymbol{k} \cdot \boldsymbol{\delta}_{ij}}  $ where $\boldsymbol{\delta}_{ij} = \boldsymbol{r}_{j} - \boldsymbol{r}_{i} $. We then transform the Hamiltonian into the symmetry basis Eq.~(\ref{sym_basis}) and impose the constraints on the parameters $t_{ij}$, such that all the symmetries of the Hamiltonian given in Section \ref{Ham_sym} are satisfied. 

To produce the band structure plot in Fig.~1(c) in the main text we include enough hopping terms in order to obtain a full matrix both in the site and the symmetry basis, hence avoiding artificial band crossings. 
The eight-band model is obtained similarly but now starting from eight sub-lattice sites (we simply add four more sub-lattice sites with $(\alpha,\beta,\gamma)\rightarrow (\alpha',\beta',\gamma')$ in Eq.~(\ref{explicit_4B})). Again, in order to produce the band structure plots in Fig.~3 in the main text, we include enough hopping terms to avoid artificial band crossings.

\section{Chern numbers using Wilson loops}

In this section we provide the full derivation of the Chern number for each Dirac point (band crossing) given in the main text. Note that the derivation is fully algebraic and thus not dependent on any particular choice of Hamiltonian. In the next section we provide a numerical calculation of the Chern numbers using a specified tight-binding Hamiltonian. 
As shown in the main text, the Chern number over a closed manifold is given by the Berry phase over its boundary.  
The total Berry phase (or abelian Wilson loop) $\gamma[\mathcal{L}]$ of a closed loop in $k$-space and over the occupied subspace $\vert \boldsymbol{u}_{occ},\boldsymbol{k} \rangle \equiv \breve{U}_{occ}(\boldsymbol{k})$ (here we take the cell periodic part of the tight-binding Bloch function) computed is given by  \cite{Bernevig1,Bernevig2,Bernevig3,Bernevig4}:
\begin{eqnarray}
	\mathrm{e}^{-i \gamma[\mathcal{L}]} &=& \mathrm{det}~ \mathcal{W}[\mathcal{L}]	\;, \\
	 \mathcal{W}[\mathcal{L}] &=& \mathrm{exp}\left[ - P \oint_{\mathcal{L}} d\boldsymbol{k} \left\langle \boldsymbol{u}_{occ},\boldsymbol{k} \right\vert \otimes \dfrac{\partial}{\partial \boldsymbol{k}} \left\vert \boldsymbol{u}_{occ},\boldsymbol{k} \right\rangle \right]  \;,
\end{eqnarray}
where $\mathcal{W}[\mathcal{L}]$ is the matrix (non-abelian) Wilson loop for the occupied subspace over $\mathcal{L}$ . Here $P$ stands for path-ordering. In the four-band subspace we have two valence bands and $\vert \boldsymbol{u}_{occ},\boldsymbol{k} \rangle = \left( \vert u_{v_1},\boldsymbol{k} \rangle,\vert u_{v_2},\boldsymbol{k} \rangle\right)^T$.
An alternative definition of the Wilson loop over a path $\mathcal{L}_{\boldsymbol{k}_2 \leftarrow \boldsymbol{k}_1}$ is given by \cite{Bernevig1,Bernevig2,Bernevig3,Bernevig4},
\begin{equation}
\label{Wilson_loop_B}
	\mathcal{W}_{\boldsymbol{k}_2 \leftarrow \boldsymbol{k}_1} =  \langle\boldsymbol{u}_{occ},\boldsymbol{k}_2  \vert  \left[ \prod\limits_{\boldsymbol{k}}^{\boldsymbol{k}_2 \leftarrow \boldsymbol{k}_1} \vert \boldsymbol{u}_{occ},\boldsymbol{k} \rangle  \langle\boldsymbol{u}_{occ},\boldsymbol{k}  \vert \right] \vert \boldsymbol{u}_{occ},\boldsymbol{k}_1 \rangle\;.
\end{equation}

\subsection{Chern number for $\mathcal{B}_{\Gamma}$}
Here we derive the Chern number for the closed manifold surrounding $\mathcal{B}_{\Gamma}$.
After the symmetry reduction based on $C_{2x}$ described in the main text, we start with $\mathcal{L} = \partial\mathcal{S}_a$. We decompose this loop into segments, each with the high-symmetry endpoints illustrated in Fig.~\ref{fig_SG19_TOP_SM}(a) (Fig.~2(b) of the main text), i.e.~we have the factorization of the Wilson loop 
\begin{equation}
\label{Wilson_a}
\mathcal{W}[\partial\mathcal{S}_a] = \mathcal{W}_{X_1\leftarrow Y_1}  \mathcal{W}_{Y_1\leftarrow T_1}  \mathcal{W}_{T_1\leftarrow U_1}  \mathcal{W}_{U_1\leftarrow T_2}\mathcal{W}_{T_2\leftarrow Y_2}\mathcal{W}_{Y_2\leftarrow X_1}\;.
\end{equation} 
\begin{figure}[t]
\centering
\begin{tabular}{cc} 
	\includegraphics[width=0.25\linewidth]{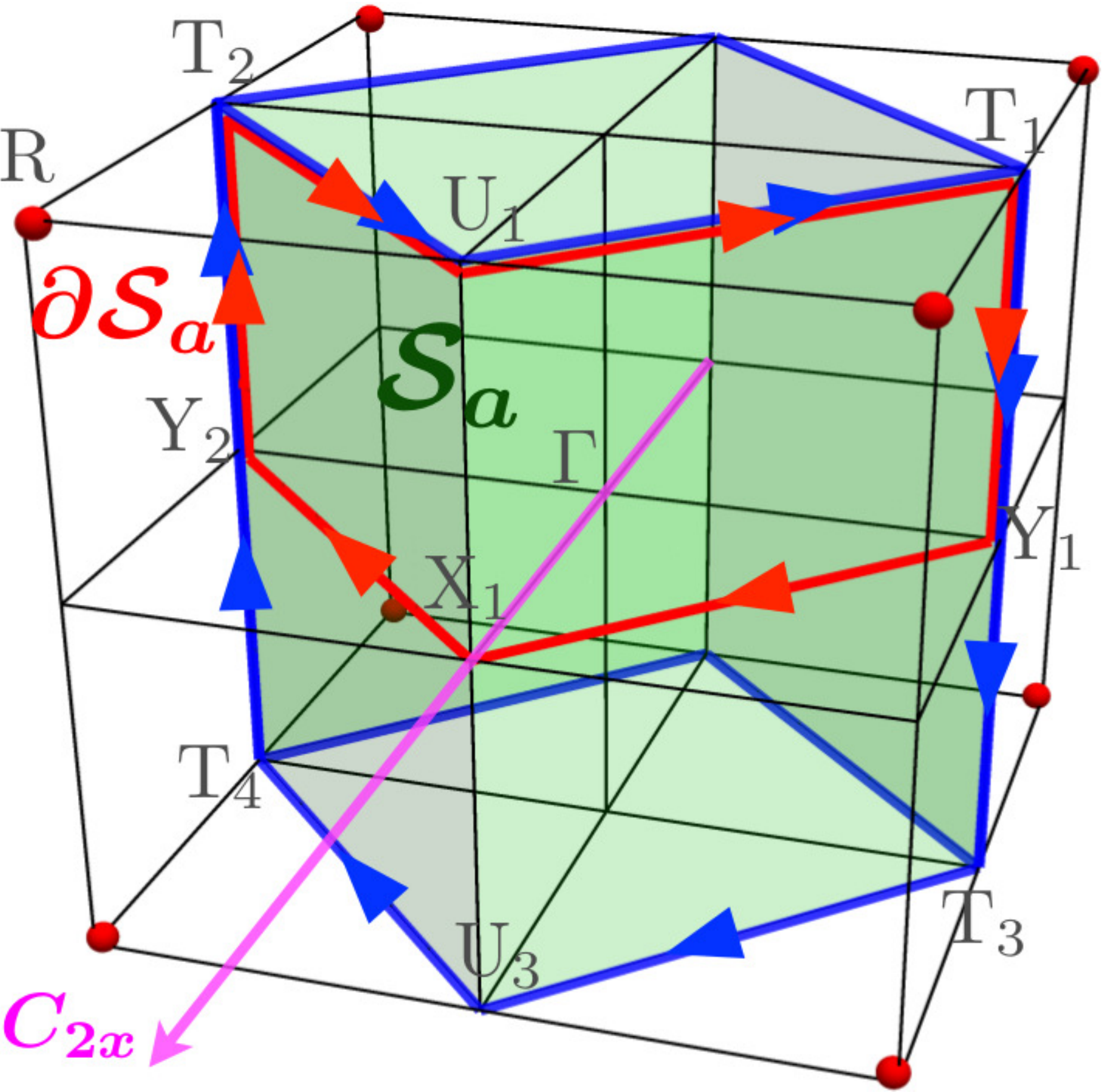} &
	\includegraphics[width=0.25\linewidth]{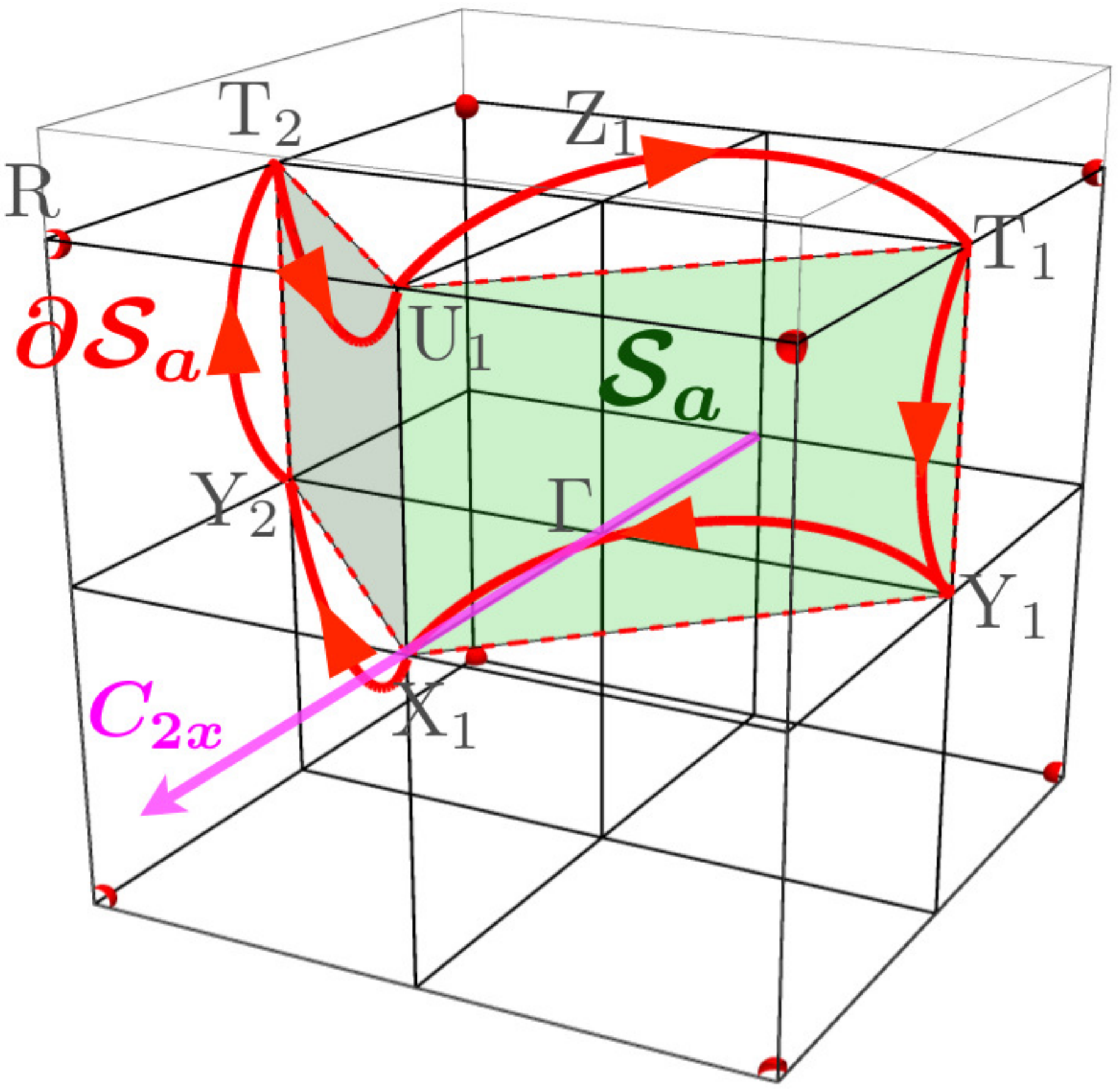} \\
	(a) & (b) 
\end{tabular}
\caption{\label{fig_SG19_TOP_SM} 
(a) Closed surface $\mathcal{S}$ separating the subspaces $\mathcal{B}_{\Gamma}$ and $\mathcal{B}_{R}$ and with the  oriented boundary $\partial \mathcal{S}_a$ of a subset $\mathcal{S}_a$, where $\mathcal{S} = \cup_{g\in D_2} g \mathcal{S}_a$. (b) Smoothly deformed oriented boundary $\partial \mathcal{S}_a$ satisfying all symmetry requirements such that its Berry phase is equal that of the boundary in (a).
} 
\end{figure}
We then use the relations, following the technique developed in Ref.~\cite{Bernevig4}:
\begin{equation}
\begin{aligned}
	\mathcal{W}_{X_1\leftarrow Y_1} &= \breve{S}^{X_1}_{occ}(\{C_{2x}\vert \boldsymbol{\tau}_x\}) \cdot \mathcal{W}_{X_1\leftarrow Y_2} \cdot \left(\breve{S}^{Y_1}_{occ}(\{C_{2x}\vert \boldsymbol{\tau}_x\})\right)^{-1} \;,\\
	\mathcal{W}_{T_1\leftarrow U_1} &=  \breve{S}^{T_4}_{occ}(\{C_{2x}\vert \boldsymbol{\tau}_x\}) \cdot\mathcal{W}_{T_4\leftarrow U_3} \cdot \left(\breve{S}^{U_1}_{occ}(\{C_{2x}\vert \boldsymbol{\tau}_x\})\right)^{-1} \;,
\end{aligned}
\end{equation}
where $ \breve{S}^{\boldsymbol{k}^*}_{occ}(\{g\vert \boldsymbol{\tau}_g\})   \equiv \langle \boldsymbol{\psi}_{occ},    g\boldsymbol{k}^* \vert ^{\{g\vert \boldsymbol{\tau}_g\}} \vert \boldsymbol{\psi}_{occ},    \boldsymbol{k}^* \rangle$.
With this Eq.~(\ref{Wilson_a}) reduces to
\begin{multline}
	\mathcal{W}[\partial\mathcal{S}_a] = \breve{S}^{X_1}_{occ}(\{C_{2x}\vert \boldsymbol{\tau}_x\})  \mathcal{W}_{X_1\leftarrow Y_2}  \left.\breve{S}^{Y_1}_{occ}(\{C_{2x}\vert \boldsymbol{\tau}_x\})\right.^{-1}  \mathcal{W}_{Y_1\leftarrow T_1} \\
	  \breve{S}^{T_4}_{occ}(\{C_{2x}\vert \boldsymbol{\tau}_x\}) \mathcal{W}_{T_4\leftarrow U_3}  \left.\breve{S}^{U_1}_{occ}(\{C_{2x}\vert \boldsymbol{\tau}_x\})\right.^{-1} 
	  \mathcal{W}_{U_1\leftarrow T_2}\mathcal{W}_{T_2\leftarrow Y_2}\mathcal{W}_{Y_2\leftarrow X_1}	\;.
\end{multline}

While we first assumed the smooth gauge in order to motivate the reduction of the Chern number over the whole $\mathcal{S}$ to the Berry phase over the boundary of the subset $\mathcal{S}_a$, we now switch to the periodic gauge. This is a mere computational trick that simplifies the symmetry reduction of the Wilson loop, as we will see, and the end result does not depend on these choices since Chern number is gauge invariant. 
On the one hand, from Eq.~(\ref{Wilson_loop_B}), assuming the periodic gauge, and by using Eq.~(\ref{band_basis_K}) we have
\begin{equation}
\label{Wilson_b}
	\mathcal{W}_{\boldsymbol{k}^*+\boldsymbol{K}_2 \leftarrow \boldsymbol{k}^* + \boldsymbol{K}_1} = \breve{M}^{\dagger}_{\boldsymbol{k}^*}(\boldsymbol{K}_2)\cdot \mathcal{W}_{\boldsymbol{k}^* \leftarrow \boldsymbol{k}^* } \cdot \breve{M}_{\boldsymbol{k}^*}(\boldsymbol{K_1}) = \mathcal{W}_{\boldsymbol{k}^* \leftarrow \boldsymbol{k}^* } \;,
\end{equation}
since $\breve{M}_{\boldsymbol{k}^*}(\boldsymbol{K}) = \boldsymbol{1}$ in the periodic gauge according to Eq.~(\ref{K_shift_band_1}) and (\ref{K_shift_band_2}). On the other hand, we have $\mathcal{W}_{\boldsymbol{k}_1 \leftarrow \boldsymbol{k}_2} = \mathcal{W}^{-1}_{\boldsymbol{k}_2 \leftarrow \boldsymbol{k}_1}$ \cite{Bernevig1,Bernevig2,Bernevig3,Bernevig4}, i.e. %
\begin{equation}
	\mathcal{W}_{\boldsymbol{k}_2 \leftarrow \boldsymbol{k}_1} \cdot \mathcal{W}_{\boldsymbol{k}_1 \leftarrow \boldsymbol{k}_2} = \boldsymbol{1} \;.
\end{equation}
Then by taking the determinant, we can reshuffle the terms and Eq.~(\ref{Wilson_b}) further reduces to
\begin{equation}
\begin{aligned}
	\det \mathcal{W}[\partial\mathcal{S}_a]& = \det  \left[ \breve{S}^{X_1}_{occ}(\{C_{2x}\vert \boldsymbol{\tau}_x\})  \left.\breve{S}^{Y_1}_{occ}(\{C_{2x}\vert \boldsymbol{\tau}_x\})\right.^{-1} 
	  \breve{S}^{T_4}_{occ}(\{C_{2x}\vert \boldsymbol{\tau}_x\}) \left.\breve{S}^{U_1}_{occ}(\{C_{2x}\vert \boldsymbol{\tau}_x\})\right.^{-1}  \right]\\
	  &= \prod\limits_{n} \dfrac{\lambda^{X}_n(\{C_{2x}\vert \boldsymbol{\tau}_x\}) \lambda^{T}_n(\{C_{2x}\vert \boldsymbol{\tau}_x\})}{\lambda^{Y}_n(\{C_{2x}\vert \boldsymbol{\tau}_x\}) \lambda^{U}_n(\{C_{2x}\vert \boldsymbol{\tau}_x\})}\;,
\end{aligned}
\end{equation}
where $\lambda^{\boldsymbol{k}^*}_n(\{C_{2x}\vert \boldsymbol{\tau}_x\}) = \lambda^{\boldsymbol{k}^*}_{x,n}$ are the eigenvalues of $\breve{S}^{\boldsymbol{k}^*}_{occ}(\{C_{2x}\vert \boldsymbol{\tau}_x\})$. This proves Eq.~(1) in the main text.
The above derivation holds for any smooth deformation of $\mathcal{S}$ and $\partial \mathcal{S}_a$, such that $\mathcal{S} = \bigcup\limits_{g\in D_2} g\mathcal{S}_a$ remains closed and such that the following relations are conserved:
\begin{equation}
\begin{aligned}
	\mathcal{L}_{Y_1 \leftarrow T_1}&= \mathcal{L}_{Y_2 \leftarrow T_2}+\boldsymbol{K}_y 	\;,\\
	C_{2x} \mathcal{L}_{X_1 \leftarrow Y_1} &= \mathcal{L}_{X_1 \leftarrow Y_2} \;,\\
	C_{2x} \mathcal{L}_{T_1 \leftarrow U_1} & = \mathcal{L}_{T_2 \leftarrow U_1} - \boldsymbol{K}_z\;.
\end{aligned}
\end{equation}
One example of an allowed deformation is given in Fig.~\ref{fig_SG19_TOP_SM}(b).

\subsection{Chern numbers for $h\overline{\Gamma X_i}$ and $\overline{X_i}$}
\begin{figure}[t]
\centering
\begin{tabular}{cc} 
	\includegraphics[width=0.25\linewidth]{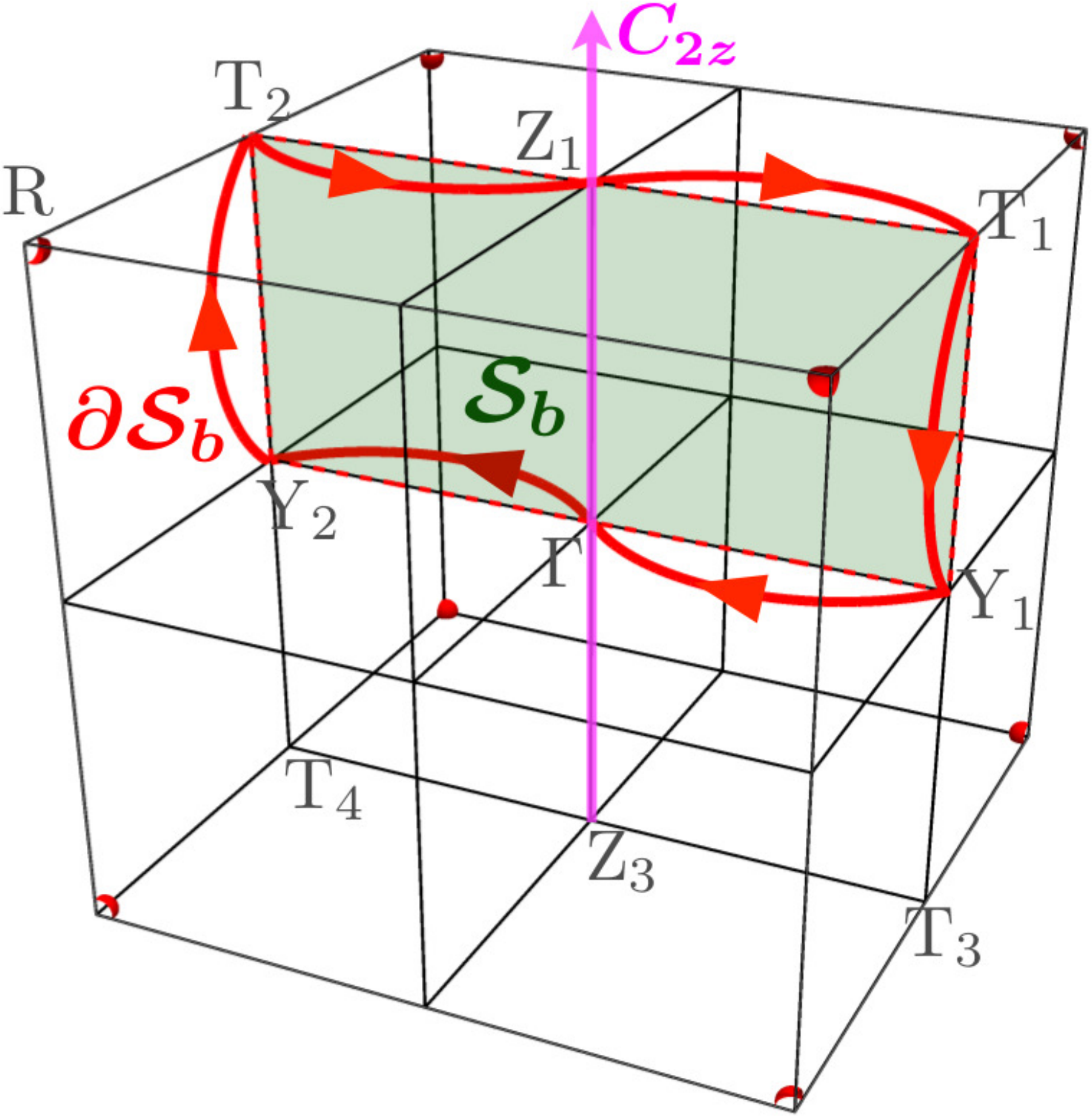} &
	\includegraphics[width=0.25\linewidth]{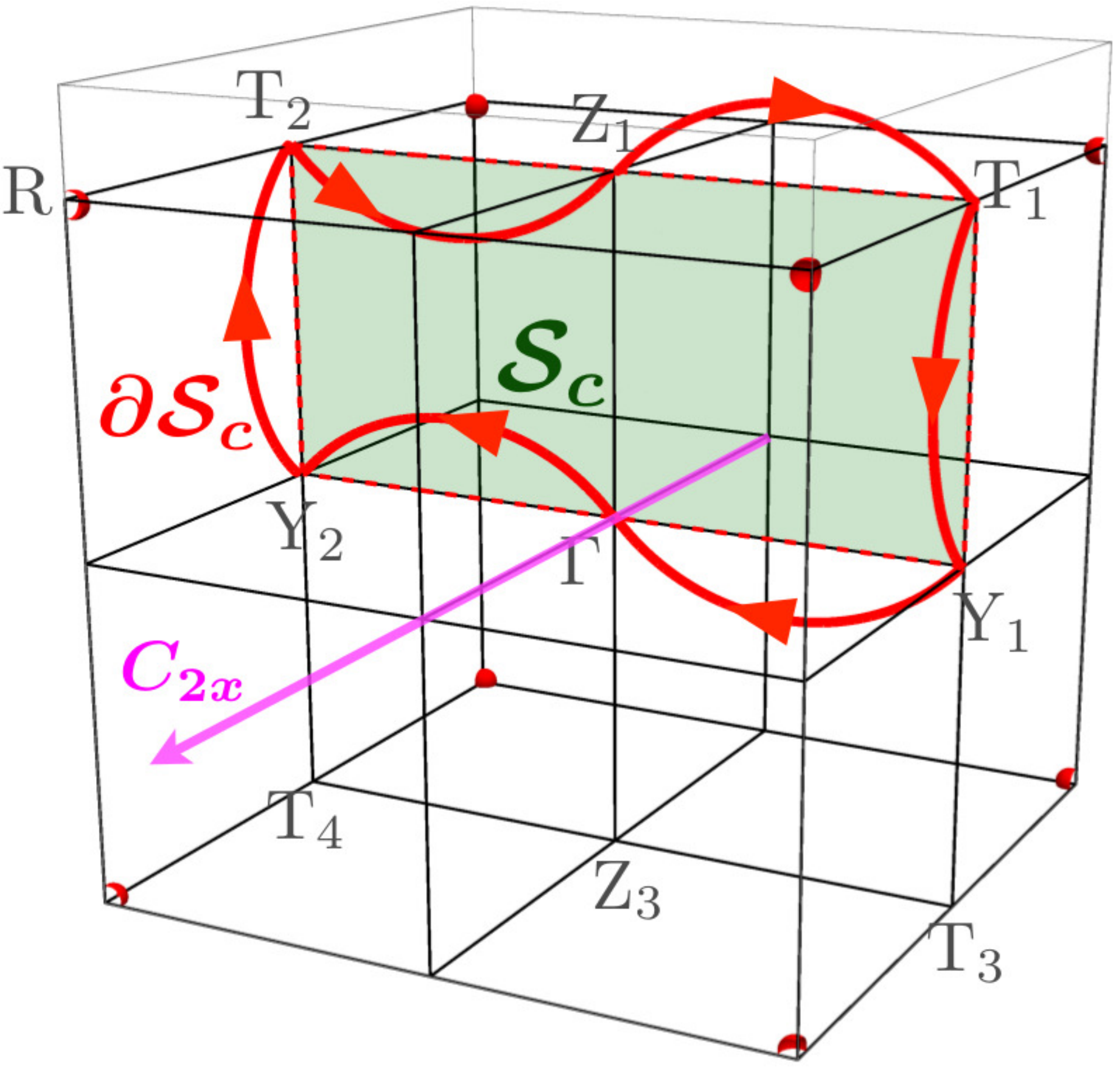} \\
	(a) & (b)
\end{tabular} 
\caption{\label{fig_SG19_TOP_SM_b} (a) Oriented boundary $\partial  \mathcal{S}_b$ of the closed surface $\mathcal{S}' = \mathcal{S}_b + C_{2z} \mathcal{S}_b$ surrounding $h\overline{\Gamma Z}$. (b) Oriented boundary $\partial  \mathcal{S}_c$ of the closed surface $\mathcal{S}'' =  \cup_{g\in D_2} g \mathcal{S}_c$ surrounding the plane $\overline{X}$. $\mathcal{S}_{(b,c)}$ is obtained from the green plane through a smooth inflation out of plane with the oriented boundary $\partial \mathcal{S}_{(b,c)}$ constrained by the symmetry requirement that $\mathcal{S}_{(b,c)} + C_{2(z,x)} \mathcal{S}_{(b,c)}$ is closed.
} 
\end{figure}
In order to differentiate between the different Dirac points within $\mathcal{B}_\Gamma$ we need to consider smaller closed surfaces, or tighter boxes, in the BZ. We start by considering $\mathcal{S}'=\mathcal{S}_b+C_{2z} \mathcal{S}_b$ and the path $\partial \mathcal{S}_b$ surrounding the half-line $h\overline{\Gamma Z}$, as exemplified in Fig.~\ref{fig_SG19_TOP_SM_b}(a) (same as Fig.~2(c) in the main text). Following similar steps as above, again assuming the periodic gauge, we have 
\begin{equation}
\begin{aligned}
	\det \mathcal{W}[\partial \mathcal{S}_b] &= \det \left[\mathcal{W}_{\Gamma\leftarrow Y_1}  \mathcal{W}_{Y_1\leftarrow T_1}  \mathcal{W}_{T_1\leftarrow Z_1}  \mathcal{W}_{Z_1\leftarrow T_2}\mathcal{W}_{T_2\leftarrow Y_2}\mathcal{W}_{Y_2\leftarrow \Gamma}\right]\\
	&= \det \left[\breve{S}^{\Gamma}_{occ}(\{C_{2z}\vert \boldsymbol{\tau}_z\}) \mathcal{W}_{\Gamma\leftarrow Y_2}  \breve{S}^{Y_1}_{occ}(\{C_{2z}\vert \boldsymbol{\tau}_z\})^{-1}\mathcal{W}_{Y_1\leftarrow T_1}  
	\breve{S}^{T_2}_{occ}(\{C_{2z}\vert \boldsymbol{\tau}_z\}) \mathcal{W}_{T_2\leftarrow Z_1}  \breve{S}^{Z_1}_{occ}(\{C_{2z}\vert \boldsymbol{\tau}_z\})^{-1}\right.\\ &\quad\quad\quad\quad\left.
	\mathcal{W}_{Z_1\leftarrow T_2}\mathcal{W}_{T_2\leftarrow Y_2}\mathcal{W}_{Y_2\leftarrow \Gamma}\right]\\
	&= \det \left[\breve{S}^{\Gamma}_{occ}(\{C_{2z}\vert \boldsymbol{\tau}_z\}) \breve{S}^{Y_1}_{occ}(\{C_{2z}\vert \boldsymbol{\tau}_z\})^{-1}  \breve{S}^{T_2}_{occ}(\{C_{2z}\vert \boldsymbol{\tau}_z\})  \breve{S}^{Z_1}_{occ}(\{C_{2z}\vert \boldsymbol{\tau}_z\})^{-1} \right]\\
	&= \prod\limits_{n} \dfrac{\lambda^{\Gamma}_n(\{C_{2z}\vert \boldsymbol{\tau}_z\}) \lambda^{T}_n(\{C_{2z}\vert \boldsymbol{\tau}_z\})  }{\lambda^{Y}_n(\{C_{2z}\vert \boldsymbol{\tau}_z\}) \lambda^{Z}_n(\{C_{2z}\vert \boldsymbol{\tau}_z\}) } \;.
\end{aligned}
\end{equation}
Using the tabulated IRREPs from Refs.~\cite{BradCrack} or \cite{Bilbao}, we finally find 
\begin{equation}
\begin{aligned}
	\mathrm{e}^{-i \pi C_1 [h\overline{\Gamma Z}]} = \mathrm{e}^{-i \gamma [\partial  \mathcal{S}_b]} &= \det \mathcal{W}[\partial \mathcal{S}_b] \\
	&=  - \lambda^{\Gamma}_{v_1}(\{C_{2z}\vert \boldsymbol{\tau}_z\}) \lambda^{\Gamma}_{v_2}(\{C_{2z}\vert \boldsymbol{\tau}_z\})\\
	&= - \chi_{z}^{v_1}\chi_{z}^{v_2}	\;,
\end{aligned}
\end{equation}
where $\chi_{i}^{v_n} = \lambda^{\Gamma}_{v_n}(\{C_{2i}\vert \boldsymbol{\tau}_i\})  $ is the character of the 1D IRREP at $\Gamma$ of the valence band $v_n$. 
The above derivation holds for any smooth deformation of $\mathcal{S}'=\mathcal{S}_b+C_{2z} \mathcal{S}_b$ and $\partial \mathcal{S}_b$, such that $\mathcal{S}'$ remains closed and that the following relations are conserved:
\begin{equation}
\begin{aligned}
	C_{2z}\mathcal{L}_{\Gamma \leftarrow Y_1}&= \mathcal{L}_{\Gamma \leftarrow Y_2} 	\;,\\
	C_{2z} \mathcal{L}_{T_1 \leftarrow Z_1} &= \mathcal{L}_{T_2 \leftarrow Z_1} \;,\\
	 \mathcal{L}_{Y_1 \leftarrow T_1} & = \mathcal{L}_{Y_2 \leftarrow T_2} + \boldsymbol{K}_y\;.
\end{aligned}
\end{equation}
The path chosen in Fig.~\ref{fig_SG19_TOP_SM_b}(a) satisfies those requirements. 
Proceeding similarly for the two other half high-symmetry lines $h\overline{\Gamma X_i}$ of $\mathcal{B}_{\Gamma}$ proves Eq.~(2) in the main text.

Next we consider $ \mathcal{S}'' = \bigcup\limits_{g \in D_2} g\mathcal{S}_c$ and the path $\partial \mathcal{S}_c$ which surrounds $\overline{X_i}$, the plane containing $\Gamma$ and perpendicular to the line $\overline{\Gamma X}$ as illustrated in Fig.~\ref{fig_SG19_TOP_SM_b}(b). Following similar steps as above, again assuming the periodic gauge, we have 
\begin{equation}
\begin{aligned}
	\det \mathcal{W}[\partial \mathcal{S}_c] &= \det \left[\mathcal{W}_{\Gamma\leftarrow Y_1}  \mathcal{W}_{Y_1\leftarrow T_1}  \mathcal{W}_{T_1\leftarrow Z_1}  \mathcal{W}_{Z_1\leftarrow T_2}\mathcal{W}_{T_2\leftarrow Y_2}\mathcal{W}_{Y_2\leftarrow \Gamma}\right]\\
	&= \det \left[\breve{S}^{\Gamma}_{occ}(\{C_{2x}\vert \boldsymbol{\tau}_x\}) \mathcal{W}_{\Gamma\leftarrow Y_2}  \breve{S}^{Y_1}_{occ}(\{C_{2x}\vert \boldsymbol{\tau}_x\})^{-1}\mathcal{W}_{Y_1\leftarrow T_1}  
	\breve{S}^{T_2}_{occ}(\{C_{2x}\vert \boldsymbol{\tau}_x\}) \mathcal{W}_{T_2\leftarrow Z_1}  \breve{S}^{Z_1}_{occ}(\{C_{2x}\vert \boldsymbol{\tau}_x\})^{-1}\right.\\ &\quad\quad\quad\quad\left.
	\mathcal{W}_{Z_1\leftarrow T_2}\mathcal{W}_{T_2\leftarrow Y_2}\mathcal{W}_{Y_2\leftarrow \Gamma}\right]\\
	&= \det \left[\breve{S}^{\Gamma}_{occ}(\{C_{2x}\vert \boldsymbol{\tau}_x\}) \breve{S}^{Y_1}_{occ}(\{C_{2x}\vert \boldsymbol{\tau}_x\})^{-1}  \breve{S}^{T_2}_{occ}(\{C_{2x}\vert \boldsymbol{\tau}_x\})  \breve{S}^{Z_1}_{occ}(\{C_{2x}\vert \boldsymbol{\tau}_x\})^{-1} \right]\\
	&= \prod\limits_{n} \dfrac{\lambda^{\Gamma}_n(\{C_{2x}\vert \boldsymbol{\tau}_x\}) \lambda^{T}_n(\{C_{2x}\vert \boldsymbol{\tau}_x\})  }{\lambda^{Y}_n(\{C_{2x}\vert \boldsymbol{\tau}_x\}) \lambda^{Z}_n(\{C_{2x}\vert \boldsymbol{\tau}_x\}) } \;.
\end{aligned}
\end{equation}
Using the tabulated IRREPs from Refs.~\cite{BradCrack} or \cite{Bilbao}, we finally find 
\begin{equation}
\begin{aligned}
	\mathrm{e}^{-i \pi \frac{C_1 [\overline{X}]}{2}} = \mathrm{e}^{-i \gamma [\partial  \mathcal{S}_c]} &= \det \mathcal{W}[\partial \mathcal{S}_c] \\
	&=  + \lambda^{\Gamma}_{v_1}(\{C_{2x}\vert \boldsymbol{\tau}_x\}) \lambda^{\Gamma}_{v_2}(\{C_{2x}\vert \boldsymbol{\tau}_x\})\\
	&= + \chi_{x}^{v_1}\chi_{x}^{v_2}	\;.
\end{aligned}
\end{equation}
The above derivation holds for any smooth deformation of $\mathcal{S}''$ and $\partial \mathcal{S}_c$, such that $\mathcal{S}_c+C_{2x} \mathcal{S}_c$ remains closed and that the following relations are conserved:
\begin{equation}
\begin{aligned}
	C_{2x}\mathcal{L}_{\Gamma \leftarrow Y_1}&= \mathcal{L}_{\Gamma \leftarrow Y_2} 	\;,\\
	C_{2x} \mathcal{L}_{T_1 \leftarrow Z_1} &= \mathcal{L}_{T_4 \leftarrow Z_3} = \mathcal{L}_{T_2 \leftarrow Z_1} -  \boldsymbol{K}_z  \;,\\
	\mathcal{L}_{Y_1 \leftarrow T_1} & = \mathcal{L}_{Y_2 \leftarrow T_2} + \boldsymbol{K}_y\;.
\end{aligned}
\end{equation}
The path chosen in Fig.~\ref{fig_SG19_TOP_SM_b}(b) satisfies those requirements. 
Proceeding similarly for the two other planes $\overline{X_i}$ crossing $\Gamma$ and perpendicular to the axes $\overline{\Gamma X_i}$ proves Eq.~(3) in the main text.

Equations~(4) and (5) in the main text for the eight-band topology are readily obtained from the above results since the only change is the doubling of the valence bands.

\subsection{Chern numbers for $h\overline{R U_i}$ and $\overline{U_i}$}

In the eight-band case we also need to characterize the topology of $\mathcal{B}_R$. This we can also do by studying appropriately chosen subdomains. We start by investigating subdomains that contain the half-lines $\{h\overline{RU_i}\}_{U_i=S,T,U}$ of the BZ boundary. 
Let us take $h\overline{RS}$ and use a surface $\mathcal{S}''' = \mathcal{S}_d + C_{2z}\mathcal{S}_d $ that surrounds it with the boundary $\partial \mathcal{S}_d = S_1 \leftarrow X_5 \leftarrow U_5 \leftarrow R \leftarrow U_1 \leftarrow X_1 \leftarrow S_1$, see Fig.~\ref{fig_SG19_TOP_SM}(a) with $X_5 = X_1 + \boldsymbol{b}_2$ and $U_5 = U_1 + \boldsymbol{b}_2$. We have 
\begin{equation}
\begin{aligned}
	\det \mathcal{W}[\partial \mathcal{S}_d] &= \det \left[\mathcal{W}_{S_1\leftarrow X_5}  \mathcal{W}_{X_5\leftarrow U_5}  \mathcal{W}_{U_5\leftarrow R}  \mathcal{W}_{R\leftarrow U_1}\mathcal{W}_{U_1\leftarrow X_1}\mathcal{W}_{X_1\leftarrow S_1}\right]\\
	&= \det \left[\breve{S}^{S_1}_{occ}(\{C_{2z}\vert \boldsymbol{\tau}_z\}) \mathcal{W}_{S_1\leftarrow X_1}  \breve{S}^{X_5}_{occ}(\{C_{2z}\vert \boldsymbol{\tau}_z\})^{-1} 
	\mathcal{W}_{X_5\leftarrow U_5}  
	\breve{S}^{U_1}_{occ}(\{C_{2z}\vert \boldsymbol{\tau}_z\}) \mathcal{W}_{U_1\leftarrow R}  \breve{S}^{R}_{occ}(\{C_{2z}\vert \boldsymbol{\tau}_z\})^{-1}\right.\\ &\quad\quad\quad\quad\left.
	\mathcal{W}_{R\leftarrow U_1}\mathcal{W}_{U_1\leftarrow X_1}\mathcal{W}_{X_1\leftarrow S_1}  \right]\\
	&= \det \left[\breve{S}^{S_1}_{occ}(\{C_{2z}\vert \boldsymbol{\tau}_z\}) \breve{S}^{X_5}_{occ}(\{C_{2z}\vert \boldsymbol{\tau}_z\})^{-1} \breve{S}^{U_1}_{occ}(\{C_{2z}\vert \boldsymbol{\tau}_z\})  \breve{S}^{R}_{occ}(\{C_{2z}\vert \boldsymbol{\tau}_z\})^{-1} \right]\;.
\end{aligned}
\end{equation}
Using the tabulated IRREPs from Refs.~\cite{BradCrack} or \cite{Bilbao}, we finally find 
\begin{equation}
\begin{aligned}
	\mathrm{e}^{-i \pi C_1 [h\overline{RS}]} = \mathrm{e}^{-i \gamma [\partial  \mathcal{S}_d]} &= \det \mathcal{W}[\partial \mathcal{S}_d] \\
	&=  (-1)^2 \det \left[ \Gamma^S_{v_1}(\{C_{2z}\vert \boldsymbol{\tau}_z\}) \Gamma^S_{v_2}(\{C_{2z}\vert \boldsymbol{\tau}_z\}) \right] \;.
\end{aligned}
\end{equation}
The above derivation holds for any smooth deformation of $\mathcal{S}'''$ and $\partial \mathcal{S}_d$, such $\mathcal{S}'''=\mathcal{S}_d+C_{2z} \mathcal{S}_d$ remains closed and that the following relations are conserved:
\begin{equation}
\begin{aligned}
	C_{2z}\mathcal{L}_{S_1 \leftarrow X_5}&= \mathcal{L}_{S_1 \leftarrow X_1} 	\;,\\
	C_{2z} \mathcal{L}_{U_5 \leftarrow R} &= \mathcal{L}_{U_1 \leftarrow R}  \;,\\
	\mathcal{L}_{U_5 \leftarrow X_5}&= \mathcal{L}_{X_1 \leftarrow U_1} + \boldsymbol{b}_2\;.
\end{aligned}
\end{equation}
Performing a similar calculation for the two other half high-symmetry lines $h\overline{RU_i}$, proves Eq.~(6) in the main text.

Let us also take the plane crossing $R$ and perpendicular to $\overline{RT}$, called $\overline{T}$, and use a surface $\mathcal{S}'''' = \bigcup_{g\in D_2}\mathcal{S}_e  $ that surrounds it, and such that the half $\mathcal{S}_e + C_{2x}\mathcal{S}_e$ of $\mathcal{S}''''$ is closed. The boundary is $\partial \mathcal{S}_e = S_1 \leftarrow X_5 \leftarrow U_5 \leftarrow R \leftarrow U_1 \leftarrow X_1 \leftarrow S_1$, similarly as above, with $R_3=R - \boldsymbol{b}_3$, $U_3=U_1 - \boldsymbol{b}_3$, and $U_6=U_5 - \boldsymbol{b}_3$. We have 
\begin{equation}
\begin{aligned}
	\det \mathcal{W}[\partial \mathcal{S}_e] &= \det \left[\mathcal{W}_{S_1\leftarrow X_5}  \mathcal{W}_{X_5\leftarrow U_5}  \mathcal{W}_{U_5\leftarrow R}  \mathcal{W}_{R\leftarrow U_1}\mathcal{W}_{U_1\leftarrow X_1}\mathcal{W}_{X_1\leftarrow S_1}\right]\\
	&= \det \left[\breve{S}^{S_1}_{occ}(\{C_{2x}\vert \boldsymbol{\tau}_x\}) \mathcal{W}_{S_1\leftarrow X_1}  \breve{S}^{X_5}_{occ}(\{C_{2x}\vert \boldsymbol{\tau}_x\})^{-1} 
	\mathcal{W}_{X_5\leftarrow U_5}  
	\breve{S}^{U_3}_{occ}(\{C_{2x}\vert \boldsymbol{\tau}_x\}) \mathcal{W}_{U_3\leftarrow R_3}  \breve{S}^{R}_{occ}(\{C_{2x}\vert \boldsymbol{\tau}_x\})^{-1}\right.\\ &\quad\quad\quad\quad\left.
	\mathcal{W}_{R\leftarrow U_1}\mathcal{W}_{U_1\leftarrow X_1}\mathcal{W}_{X_1\leftarrow S_1}  \right]\\
	&= \det \left[\breve{S}^{S_1}_{occ}(\{C_{2x}\vert \boldsymbol{\tau}_x\}) \breve{S}^{X_5}_{occ}(\{C_{2x}\vert \boldsymbol{\tau}_x\})^{-1} \breve{S}^{U_3}_{occ}(\{C_{2x}\vert \boldsymbol{\tau}_x\})  \breve{S}^{R}_{occ}(\{C_{2x}\vert \boldsymbol{\tau}_x\})^{-1} \right]\\
	&=  \prod\limits_{n} \dfrac{\lambda^{S}_n(\{C_{2x}\vert \boldsymbol{\tau}_x\}) \lambda^{U}_n(\{C_{2x}\vert \boldsymbol{\tau}_x\})  }{\lambda^{X}_n(\{C_{2x}\vert \boldsymbol{\tau}_x\}) \lambda^{R}_n(\{C_{2x}\vert \boldsymbol{\tau}_x\}) } \;.
\end{aligned}
\end{equation}
Using the tabulated IRREPs from Refs.~\cite{BradCrack} or \cite{Bilbao}, we finally find 
\begin{equation}
\begin{aligned}
	\mathrm{e}^{-i \pi\frac{C_1 [\overline{T}]}{2}} = \mathrm{e}^{-i \gamma [\partial  \mathcal{S}_e]} &= \det \mathcal{W}[\partial \mathcal{S}_e] \\
	&= +1 \;.
\end{aligned}
\end{equation}
The above derivation holds for any smooth deformation of $\mathcal{S}_e$ and $\partial \mathcal{S}_e$, such that $\mathcal{S}''''$ and $\mathcal{S}_e+C_{2x} \mathcal{S}_e$ remain closed and that the following relations are conserved:
\begin{equation}
\begin{aligned}
	C_{2x}\mathcal{L}_{S_1 \leftarrow X_5}&= \mathcal{L}_{S_1 \leftarrow X_1} 	\;,\\
	C_{2x} \mathcal{L}_{U_5 \leftarrow R} &= \mathcal{L}_{U_3 \leftarrow R_3} = \mathcal{L}_{U_1 \leftarrow R} -  \boldsymbol{b}_3  \;,\\
	\mathcal{L}_{U_5 \leftarrow X_5}&= \mathcal{L}_{X_1 \leftarrow U_1} + \boldsymbol{b}_2\;.
\end{aligned}
\end{equation}
Proceeding similarly for the other planes $\overline{U_i}$ proves the result $C_1[\overline{U_i}] = 0 \mod 4$ in the main text.

\section{Numerical computation of the Chern numbers}
Above we showed the existence of symmetry protected band-crossing points, or Dirac points, in any four-band system with SG19: a pair of simple Dirac points must be realized along one of the high-symmetry axes of $\mathcal{B}_{\Gamma}$ and a double Dirac point must be realized at the $R$ point. 
Here we provide a numerical calculation of the Chern number for the Dirac points present in the explicit four-band Hamiltonian used in in Fig.~1(c) of the main text. Note that ordering the eigenvalues in energy after solving a band-eigenvalue problem numerically naturally results in working within the periodic gauge. 

For this, we define a closed surface that surrounds a band-crossing point, for instance a sphere centered on the crossing point $\mathcal{S}_{\mathrm{DP}}$ as illustrated in Fig.~\ref{fig_SG19_numerical_Chern}(a). We then define a loop on the sphere at a constant polar angle $\theta$, which we write $\mathcal{L}_{\theta}$. Scanning through all polar angles, $\theta\in [0,\pi]$, we cover the whole sphere, starting at the north pole and ending at the south pole. For each loop we compute numerically the Berry phase following the technique of Refs.~\cite{Wilson_flow1,Wilson_flow2}. The Chern number is then simply the total flow of the Berry phase (modulo $2\pi$), obtained through the parallel transport of the loops over the sphere (for an early formulation of this approach, see Ref.~\cite{Stone_76}):
\begin{equation}
	2\pi C_1(\mathrm{DP}) = [\Delta \gamma]_{\mathcal{S}_{\mathrm{DP}}} = \gamma[\mathcal{L}_{\pi}]  - \gamma[\mathcal{L}_{0}]\;.
\end{equation}

We show in Fig.~\ref{fig_SG19_numerical_Chern}(b) the Berry phase flow over the sphere in Fig.~\ref{fig_SG19_numerical_Chern}(a), which surrounds a band-crossing point located on along the $\Gamma Z$ line. Since there is a total flow of $\Delta\gamma = -2\pi$, the Dirac point has a Chern number $-1$ (the sign is arbitrary), hence it acts as a sink of Berry flux. We obtain the same result for the second inequivalent Dirac point on the $k_z$-axis. Since the global charge over the BZ must cancel, we deduce that the double crossing point at $R$ must then have a Chern number of $+2$. Indeed, performing the same computation around the $R$-point we obtain the Berry phase flow shown in  Fig.~\ref{fig_SG19_numerical_Chern}(c), which leads to a Chern number $+2$. The $R$-point is a double Dirac point that acts as a source of Berry flux in the BZ.
\begin{figure}[t]
\centering
\begin{tabular}{ccc} 
	\includegraphics[width=0.33\linewidth]{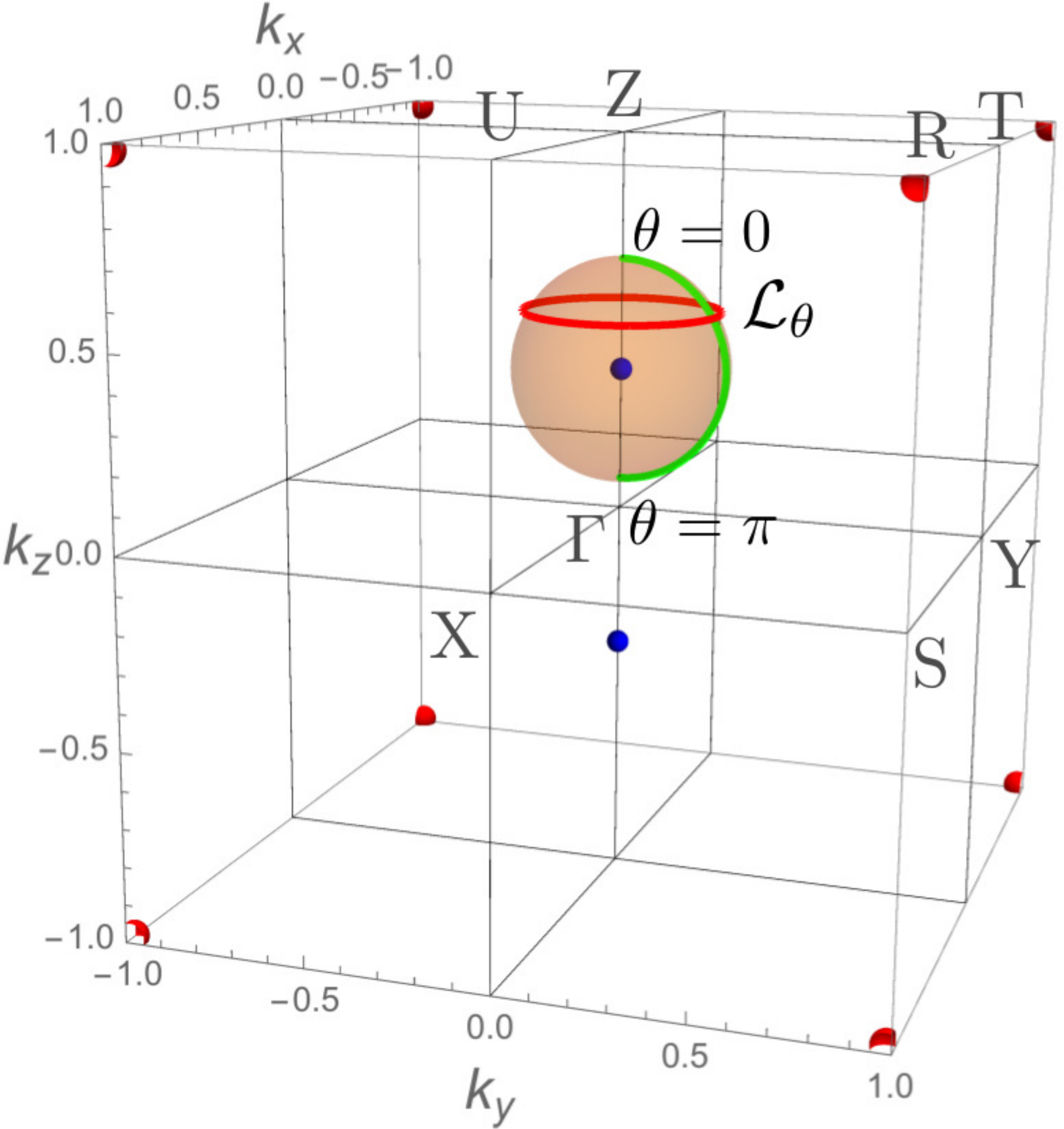} &
	\includegraphics[width=0.33\linewidth]{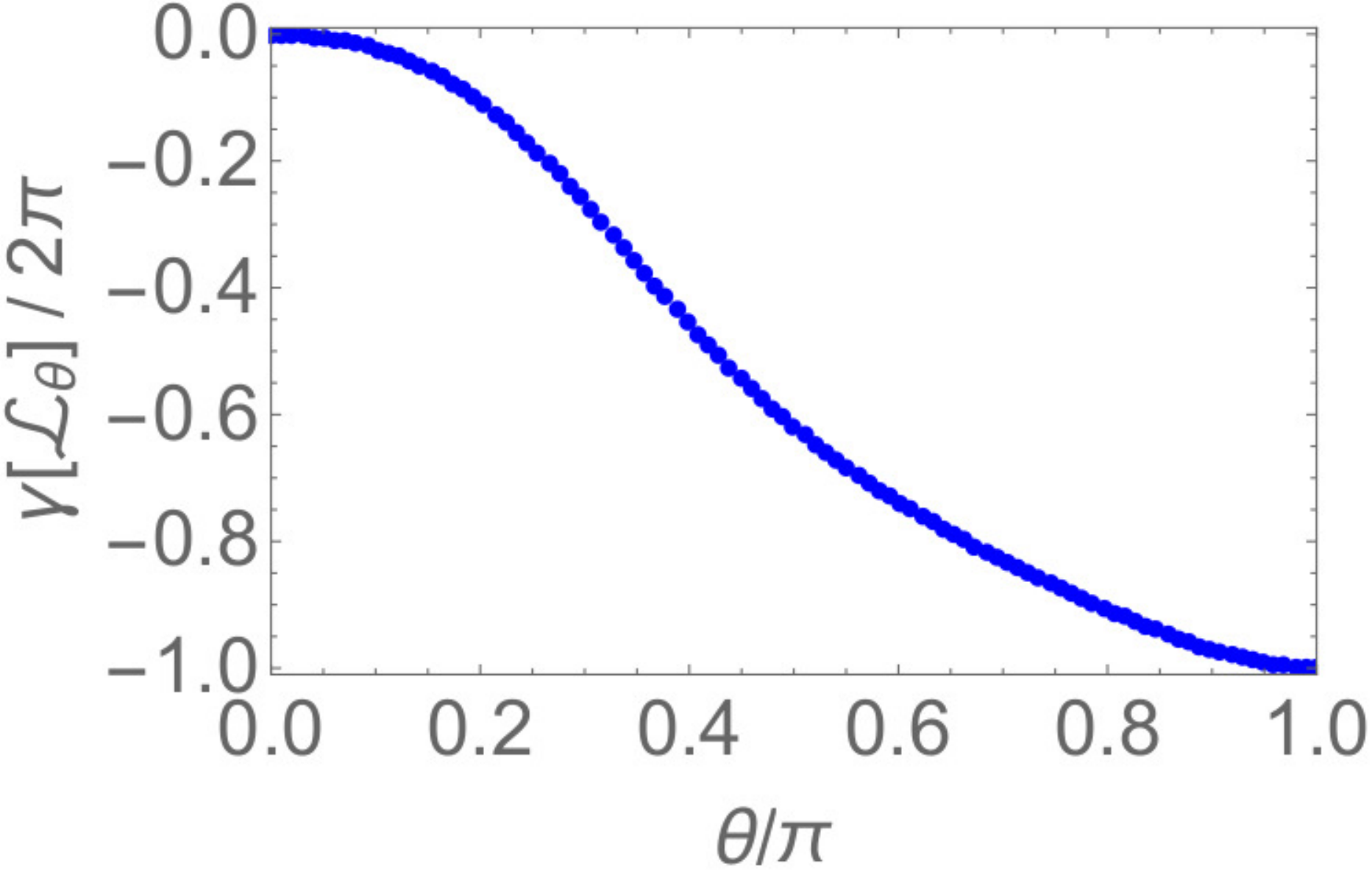}& 
	\includegraphics[width=0.33\linewidth]{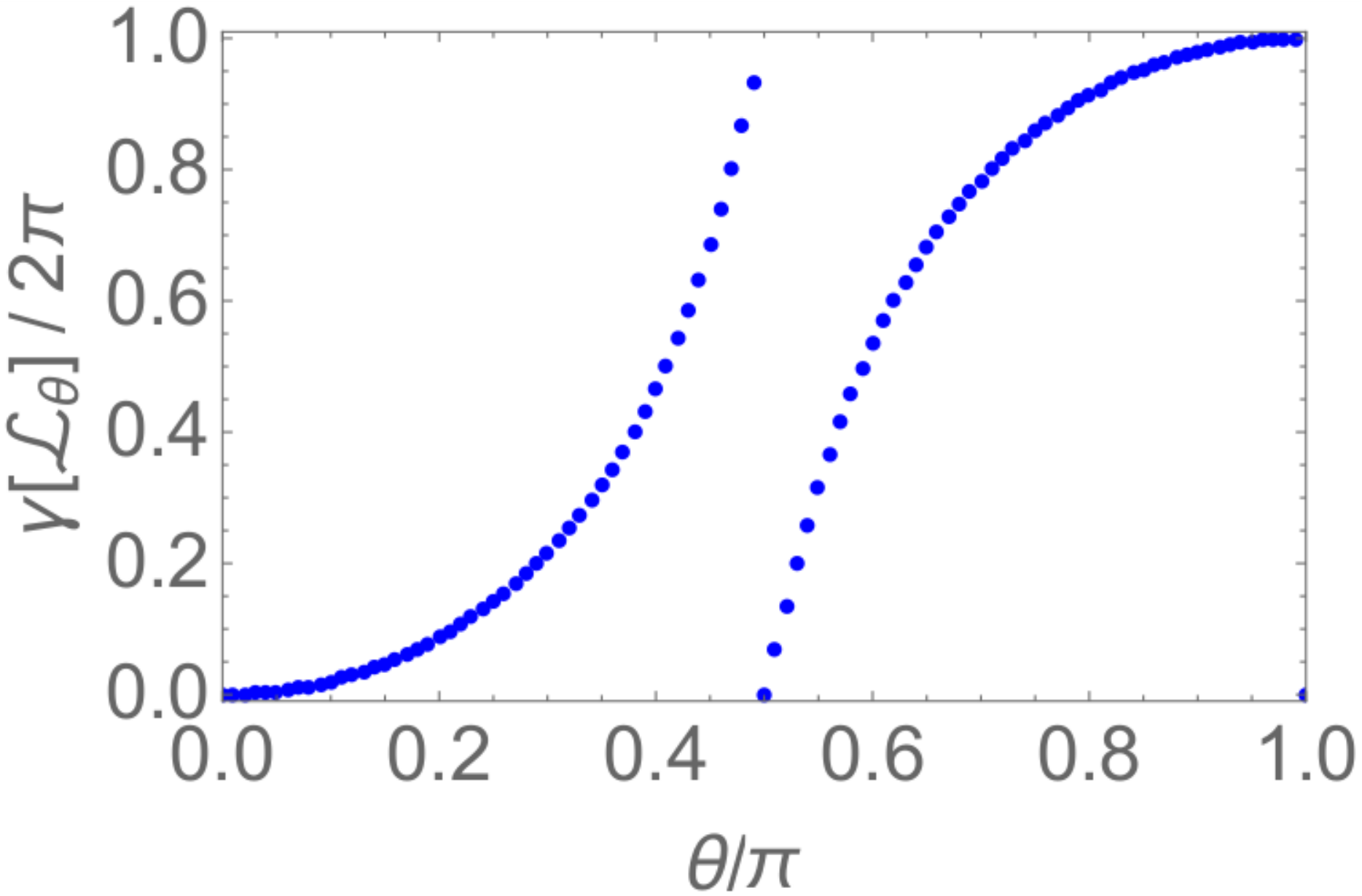}\\
	$(a)$ & $(b)$& $(c)$ 
\end{tabular}
\caption{\label{fig_SG19_numerical_Chern} $(a)$ Sphere surrounding a band-crossing point along $\Gamma Z$, with the loops $\mathcal{L}_\theta$ parametrized by the polar angle $\theta\in [0,\pi]$. $(b)$ Flow of the Berry phase $\gamma[\mathcal{L}_{\theta}]$ as we sweep the loops over the sphere in  $(a)$. (c) Flow of the Berry phase over a sphere surrounding the $R$ point.}
\end{figure}
%

\section{Breaking time reversal symmetry} 
Let us briefly also address the case when TRS is broken. In Section I we have commented on when breaking TRS changes the IRREPs of the system. The two simple Dirac points appearing due to the four-band connectivity are protected by space group symmetries only and will be present even without TRS. However, the fourfold degeneracy at $R$ requires TRS. When TRS is broken the 4D IRREP at $R$ splits into two equivalent 2D IRREPs, which further split into four nonequivalent 1D IRREPs along the high-symmetry lines $\{\overline{RS},\overline{RT},\overline{RU}\}$ \cite{BradCrack}. 
Therefore, three pairs of Dirac points are produced; one pair of Dirac points on each line $\{\overline{RS},\overline{RT},\overline{RU}\}$. Let us assume that we break TRS adiabatically, i.e.~with no band inversion at the other high-symmetry points. In that case, the total charge of the three pairs of Dirac points must be equal to the charge of the original TRS double Dirac point. However, in general, breaking TRS easily leads to the formation of quadruples of Dirac points associated with one arbitrary $\boldsymbol{k}$-point within the BZ (if one Dirac point is realized at an arbitrary $\boldsymbol{k}$-point, it must be accompanied by its three partners under the orbit $\{g\boldsymbol{k}\vert g\in D_2\}$). Thus, counting the charges of the Dirac points located at high-symmetry points is an easy way to infer the existence of other Dirac points at arbitrary $\boldsymbol{k}$-points. 

\end{widetext}

\bibliography{mybib}

\clearpage

\end{document}